\documentclass[12pt]{article}

\usepackage[nosort]{cite}
\usepackage{graphicx}
\usepackage{pst-all}
\usepackage{bbm}
\usepackage{amsmath}
\usepackage{amssymb}
\usepackage{subfigure}
\usepackage{slashed}
\usepackage{multirow}
\usepackage{arydshln} % dashed lines in tables

\makeatletter \@addtoreset{equation}{section} \makeatother

\makeatletter
\let\old@startsection=\@startsection
\let\oldl@section=\l@section
\renewcommand{\@startsection}[6]{\old@startsection{#1}{#2}{#3}{#4}{#5}{#6\mathversion{bold}}}
\renewcommand{\l@section}[2]{\oldl@section{\mathversion{bold}#1}{#2}}
\makeatother

\makeatletter
\let\old@makecaption=\@makecaption
\def\@makecaption{\small\old@makecaption}
\makeatother

\renewcommand{\thefootnote}{\arabic{footnote}}
\let\oldPhi=\Phi
\let\oldPsi=\Psi
\let\oldGamma=\Gamma
\let\oldDelta=\Delta
\let\oldSigma=\Sigma
\let\oldTheta=\Theta
\let\oldPi=\Pi
\let\oldUpsilon=\Upsilon
\renewcommand{\Phi}{\mathnormal{\oldPhi}}
\renewcommand{\Psi}{\mathnormal{\oldPsi}}
\renewcommand{\Gamma}{\mathnormal{\oldGamma}}
\renewcommand{\Sigma}{\mathnormal{\oldSigma}}
\renewcommand{\Delta}{\mathnormal{\oldDelta}}
\renewcommand{\Theta}{\mathnormal{\oldTheta}}
\renewcommand{\Pi}{\mathnormal{\oldPi}}
\renewcommand{\Upsilon}{\mathnormal{\oldUpsilon}}

\newcommand{\superN}{\mathcal{N}}
\newcommand{\Action}{\mathcal{S}}

\newcommand{\tr}{\mathop{\mathrm{tr}}}

\newcommand{\diag}{\mathop{\mathrm{diag}}}
\newcommand{\sign}{\mathop{\mathrm{sign}}}

\newcommand{\trans}{{\scriptscriptstyle\mathrm{T}}}

\newcommand{\Integers}{\mathbbm{Z}}
\newcommand{\Reals}{\mathbbm{R}}

\newcommand{\Sphere}{S}  % {\mathbbm{S}}

\ifx\genfrac\sdflkaj

\else

\fi
\newcommand{\sfrac}[2]{{\textstyle\frac{#1}{#2}}}
\newcommand{\half}{\sfrac{1}{2}}
\newcommand{\ihalf}{\sfrac{i}{2}}
\newcommand{\quarter}{\sfrac{1}{4}}

\newcommand{\Half}{\frac{1}{2}}
\newcommand{\iHalf}{\frac{i}{2}}
\newcommand{\Quarter}{\frac{1}{4}}

\newcommand{\rep}[1]{{\mathbf{#1}}}
\newcommand{\matr}[2]{\left(\begin{array}{#1}#2\end{array}\right)}

\newcommand{\grp}[1]{\mathrm{#1}}

\newcommand{\grU}{\grp{U}}
\newcommand{\grSU}{\grp{SU}}
\newcommand{\grSO}{\grp{SO}}

\newcommand{\lrbrk}[1]{\left(#1\right)}
\newcommand{\bigbrk}[1]{\bigl(#1\bigr)}
\newcommand{\Bigbrk}[1]{\Bigl(#1\Bigr)}

\newcommand{\lrsbrk}[1]{\left[#1\right]}
\newcommand{\bigsbrk}[1]{\bigl[#1\bigr]}
\newcommand{\Bigsbrk}[1]{\Bigl[#1\Bigr]}
\newcommand{\biggsbrk}[1]{\biggl[#1\biggr]}

\newcommand{\comm}[2]{[#1,#2]}

\newcommand{\acomm}[2]{\{#1,#2\}}

\newcommand{\lrabs}[1]{\left|#1\right|}
\newcommand{\abs}[1]{{|#1|}}
\newcommand{\bigabs}[1]{\bigl|#1\bigr|}

\newcommand{\nn}{\nonumber}

\newcommand{\nl}[1][0pt]{\nonumber\\[#1]&\hspace{-4\arraycolsep}&\mathord{}}

\newcommand{\earel}[1]{\mathrel{}&\hspace{-2\arraycolsep}#1\hspace{-2\arraycolsep}&\mathrel{}}
\newcommand{\eq}{\earel{=}}

\def\[{\begin{equation}}
\def\]{\end{equation}}

\makeatletter
\def\mr@ignsp#1 {\ifx\:#1\@empty\else #1\expandafter\mr@ignsp\fi}%
\newcommand{\multiref}[1]{\begingroup%\let\protect\string%
\xdef\mr@no@sparg{\expandafter\mr@ignsp#1 \: }%
\def\mr@comma{}%
\@for\mr@refs:=\mr@no@sparg\do{\mr@comma\def\mr@comma{,}\ref{\mr@refs}}%
\endgroup}
\makeatother

\newcommand{\hypref}[2]{\ifx\href\asklfhas #2\else\href{#1}{#2}\fi}

\newcommand{\secref}[1]{Sec.~\multiref{#1}}

\newcommand{\appref}[1]{App.~\multiref{#1}}

\newcommand{\tabref}[1]{Tab.~\multiref{#1}}

\newcommand{\figref}[1]{Fig.~\multiref{#1}}
\renewcommand{\eqref}[1]{(\multiref{#1})}

\ifx\href\asklfhas\newcommand{\href}[2]{#2}\fi

\newcommand{\comma}{\quad,\quad}
\newcommand{\unit}{\mathbbm{1}}

\newcommand{\Bsi}{\Upsilon}

\newcommand{\levi}{\epsilon}

\newcommand{\eps}{\varepsilon}

\newcommand{\be}{\begin{eqnarray}}
\newcommand{\ee}{\end{eqnarray}}

\newcommand{\auxD}{\mathsf{D}}
\newcommand{\deriD}{\mathcal{D}}
\newcommand{\WW}{\mathcal{W}}

\newcommand{\ZZ}{\mathcal{Z}}
\newcommand{\VV}{\mathcal{V}}

\newcommand{\superW}{\mathrm{W}}

\newcommand{\ha}{{\hat{a}}}
\newcommand{\hb}{{\hat{b}}}

\newcommand{\tg}{{\tilde{g}}}

\newcommand{\ext}{\mathrm{d}}

\newcommand{\backslashed}[1]{#1\!\!\!\!\backslash\,}

\newcommand{\lrderiD}{\overleftrightarrow{\deriD}\!}

\newcommand{\chisigma}{\chi_\sigma^{\vphantom{\dagger}}}
\newcommand{\chiphi}{\chi_\phi^{\vphantom{\dagger}}}
\newcommand{\hchisigma}{\hat{\chi}_\sigma^{\vphantom{\dagger}}}
\newcommand{\hchiphi}{\hat{\chi}_\phi^{\vphantom{\dagger}}}

\begin{document}

\thispagestyle{empty}
\begin{flushright}\footnotesize
\texttt{arXiv:0906.3008}\\
\texttt{PUPT-2304} \vspace{10mm}
\end{flushright}

\renewcommand{\thefootnote}{\fnsymbol{footnote}}
\setcounter{footnote}{0}

\begin{center}
{\Large\textbf{\mathversion{bold}
Charges of Monopole Operators \\
in Chern-Simons Yang-Mills Theory
}\par}

\vspace{1.5cm}

\textrm{Marcus K.\ Benna$^{a}$, Igor R.\ Klebanov$^{a,b}$ and Thomas Klose$^{b}$} \vspace{8mm} \\
\textit{
$^a$Joseph Henry Laboratories and $^b$Princeton Center for Theoretical Science \\
Princeton University, Princeton, NJ 08544, USA
} \\
\texttt{\\ mbenna,klebanov,tklose@princeton.edu}

%%%%%%%%
\par\vspace{14mm}

\textbf{Abstract} \vspace{5mm}

\begin{minipage}{14cm}
We calculate the non-abelian R-charges of BPS monopole operators in three-dimensional gauge theories with $\superN=3$ supersymmetry. This class of models includes ABJM theory, the proposed gauge theory dual of M-theory on $AdS_4\times S^7/\Integers_k$. In the UV limit of the $\superN=3$ theories the Yang-Mills coupling becomes weak and the monopole operators are described by classical backgrounds. This allows us to find their $\grSU(2)_R$ charges in a one-loop computation which by virtue of the non-renormalization of non-abelian R-charges yields the exact result for any value of the coupling. The spectrum of $\grSU(2)_R$ charges is found by quantizing the $\grSU(2)/\grU(1)$ collective coordinate of the BPS background, whose dynamics is that of a charged particle on a sphere with a Wess-Zumino term representing a magnetic monopole at its center. If the Wess-Zumino coefficient is $h$, then the smallest possible $\grSU(2)_R$ representation for BPS monopole operators has spin $\abs{h}/2$. We find, in agreement with earlier proposals, that $h$ is proportional to the sum of the $\grU(1)_R$ charges of all the fermion fields weighted by the effective monopole charges determined by their gauge representations. The field content of ABJM theory is such that $h=0$. This proves for any Chern-Simons level $k$ the existence of monopole operators which are singlets under all global symmetries and have vanishing scaling dimensions. These operators are essential for matching the spectrum of the ABJM theory with supergravity and for the supersymmetry enhancement to $\superN=8$.
\end{minipage}

\end{center}

\vspace{0.5cm}

%%%%%%%%%%%%%%%%%%%%%%%%%%%%%%%%%%%%%%%%%%%%%%%%%%%%%%%%%%%%%%%%%%%%%%%%%%%
\newpage
\setcounter{page}{1}
\renewcommand{\thefootnote}{\arabic{footnote}}
\setcounter{footnote}{0}

\hrule
\tableofcontents
\vspace{8mm}
\hrule
\vspace{4mm}

%%%%%%%%%%%%%%%%%%%%%%%%%%%%%%%%%%%%%%%%%%%%%%%%%%%%%%%%%%%%%%%%%%%%%%%%%%%
%%%%%%%%%%%%%%%%%%%%%%%%%%%%%%%%%%%%%%%%%%%%%%%%%%%%%%%%%%%%%%%%%%%%%%%%%%%
\section{Introduction}

Superconformal Chern-Simons gauge theories are excellent candidates for describing the dynamics of coincident M2-branes \cite{Schwarz:2004yj}. Bagger and Lambert \cite{Bagger:2006sk, Bagger:2007jr, Bagger:2007vi}, and Gustavsson \cite{Gustavsson:2007vu} succeeded in constructing the first $\superN=8$ supersymmetric classical actions for Chern-Simons gauge fields coupled to matter. Requiring manifest unitarity restricts the gauge group to $\grSO(4)$ \cite{Gauntlett:2008uf,Papadopoulos:2008sk}; this model may be reformulated as $\grSU(2)\times \grSU(2)$ gauge theory with conventional Chern-Simons terms having opposite levels $k$ and $-k$ \cite{VanRaamsdonk:2008ft,Bandres:2008vf}. Aharony, Bergman, Jafferis, and Maldacena (ABJM) \cite{Aharony:2008ug} proposed that a similar $\grU(N)\times \grU(N)$ Chern-Simons gauge theory with levels $k$ and $-k$ arises on the world volume of $N$ M2-branes placed at the singularity of $\Reals^8/\Integers_k$, where $\Integers_k$ acts by simultaneous rotation in the four planes. Therefore, the ABJM theory was conjectured to be dual, in the sense of AdS/CFT correspondence \cite{Maldacena:1997re,Gubser:1998bc,Witten:1998qj}, to M-theory on $AdS_4\times S^7/\Integers_k$. For $k>2$ this orbifold preserves only $\superN=6$ supersymmetry, and so does the ABJM theory \cite{Aharony:2008ug,Benna:2008zy,Bandres:2008ry}. The conjectured duality predicts that for $k=1,2$ the supersymmetry of the gauge theory must be enhanced to $\superN=8$. \\

The mechanism for this symmetry enhancement in the quantum theory was suggested in \cite{Aharony:2008ug}; it relies on the existence of certain monopole operators in 3-d gauge theories \cite{Hooft:1977hy,Polyakov:1976fu,Moore:1989yh,Borokhov:2002ib,Borokhov:2002cg,Borokhov:2003yu,Itzhaki:2002rc}\footnote{A more appropriate name may be ``instanton operators'' since they create instantons of a Euclidean 3-d theory whose spacetime dependence resembles the spatial profile of monopoles in 3+1 dimensions.}. Insertion of such an operator at some point creates quantized flux in a $\grU(1)$ subgroup of the gauge group through a sphere surrounding this point. For example, in a $\grU(1)$ gauge theory on $\Reals^3$, a monopole operator placed at the origin creates the Dirac monopole field
\be \label{eqn:Dirac}
  A =\frac{H}{2} \, \frac{\pm 1 - \cos\theta}{r} \, \ext\varphi
\ee
where the upper sign is for the northern hemisphere and lower sign for the southern one. In addition, some scalar fields may need to be turned on as well; they are required for BPS monopoles that preserve supersymmetry. The fluctuations of fermionic matter fields can shift the dimension of such a monopole operator. These effects were studied in some simple models, mostly with $\grU(1)$ gauge group, in \cite{Borokhov:2002ib,Borokhov:2002cg,Borokhov:2003yu}. In this paper we will generalize these calculations to more complicated models. Our primary goal is to calculate R-charges and dimensions of the monopole operators in ABJM theory for any level $k$. Since for small values of $k$ we cannot use perturbation theory in $1/k$, we will actually study the $\superN=3$ supersymmetric Yang-Mills Chern-Simons theory that provides a weakly coupled UV completion of the ABJM theory.\footnote{We are grateful to Juan Maldacena for this suggestion. A similar trick was used in \cite{Kim:2009wb}.} \\

In gauge theories with $\grU(N)$ gauge group there exists a rich set of monopole operators labeled by the generator $H$ that specifies the embedding of $\grU(1)$ into $\grU(N)$ \cite{Kapustin:2006pk} (see appendix D of \cite{Klebanov:2008vq} for a brief discussion). The generalized Dirac quantization condition restricts, up to gauge equivalence, the background gauge field to be proportional to the Cartan generator $H = \diag(q_1, \ldots, q_N)$ where the integers $q_i$ satisfy $q_1\geq q_2 \ldots \geq q_N$. If the action of the gauge theory includes a Chern-Simons term with level $k$, then the monopole operators are expected to transform non-trivially under the $\grU(N)$ gauge group, in $\grU(N)$ representations given by the Young tableaux with rows of length $kq_1, kq_2, \ldots, kq_N$ \cite{Kapustin:2006pk}. \\

The monopole operators in the $\grU(N)\times \grU(N)$ ABJM theory have been a subject of several recent investigations \cite{Berenstein:2008dc,Klebanov:2008vq,Imamura:2009ur,Kim:2009wb,SheikhJabbari:2009kr}. In general the monopoles are described by two different generators, $H$ and $\hat{H}$, which specify the form of the two gauge potentials, $A$ and $\hat A$, subject to the constraint $\sum_i q_i = \sum_i \hat{q}_i$ \cite{Kim:2009wb}. 
%However, the monopoles are believed to be BPS and thus not suffer renormalization of their dimensions only for 
We will mostly restrict our attention to the BPS monopoles with $H=\hat H$. A proposal for the $\grU(1)$ R-charge of the monopole operator in gauge theories with $\superN=3$ supersymmetry was made in \cite{Gaiotto:2008ak,Gaiotto:2009tk} based on the results of \cite{Borokhov:2002cg} and group theoretic arguments. It states that the R-charge induced by fermionic fluctuations is
\be \label{eqn:induced}
Q_R^{\mathrm{mon}} = \Half \lrbrk{ \sum_{i} \abs{h_i} - \sum_{j} \abs{v_i} }
\ee
where $h_i$ and $v_j$ are, respectively, the R-charges of the fermions in hyper and vector multiplets weighted by the effective monopole charge appropriate for their gauge representations. We establish this formula through an explicit calculation in \secref{sec:U1}, and derive its non-abelian generalization in \secref{sec:SU2}. In the ABJM theory, if we consider the diagonal $\grU(N)$, we find two hyper multiplets and two vector multiplets with charges such that there is a desired cancellation of the anomalous dimension. Using the monopole operators $(\mathcal{M}^{-2})_{ab}^{\ha\hb}$ with $kq_1=k\hat{q}_1=2$ which exist for $k=1,2$, we can form twelve conserved currents of dimension $2$,
\be
  J_\mu^{AB} =  \mathcal{M}^{-2} \Bigsbrk{ Y^A \deriD_\mu Y^B - \deriD_\mu Y^A Y^B + i \psi^{\dagger A} \gamma^\mu \psi^{\dagger B} }
\ee
and their complex conjugates, which are responsible for the symmetry enhancement from $\grSU(4)_R$ to $\grSO(8)_R$. \\

Monopoles are crucial not only for the supersymmetry enhancement at level $k=1,2$ but also for matching the spectrum of the dual gravity theory at any level $k$. In fact, supergravity modes with momentum along the M-theory direction are dual to gauge invariant operators involving monopoles. The spectra match if there are monopole operators which can render a gauge theory operator gauge invariant without altering the global charges and dimensions of the matter fields. These monopole operator themselves have to be singlets under all global symmetries and have to have vanishing dimension. In $\grU(N)\times \grU(N)$ gauge theories coupled to $N_f$ bifundamental hyper multiplets, this dimension is proportional to $N_f-2$. Thus, the requisite monopole operators exist in the ABJM theory, which has $N_f=2$. \\

The outline of the paper is as follow. \secref{sec:heart} summarizes our reasoning and the results, and ties together the subsequent more technical sections. \secref{sec:CSYM} contains the details of the theories under consideration. In \secref{sec:U1} and \secref{sec:SU2} we present the computation of the $\grU(1)_R$ and $\grSU(2)_R$ charges of the monopole operators, respectively. Brief conclusions and an outlook are given in \secref{sec:conclusions}.

%%%%%%%%%%%%%%%%%%%%%%%%%%%%%%%%%%%%%%%%%%%%%%%%%%%%%%%%%%%%%%%%%%%%%%%%%%%
%%%%%%%%%%%%%%%%%%%%%%%%%%%%%%%%%%%%%%%%%%%%%%%%%%%%%%%%%%%%%%%%%%%%%%%%%%%
\section{Summary}
\label{sec:heart}

In this section we go through the logic of our arguments and present the results of our computations while omitting the technical details. \\

A monopole operator is defined by specifying the singular behavior of the gauge field, as in (\ref{eqn:Dirac}), and appropriate matter fields close to the insertion point. As such it cannot be written as a polynomial in the fundamental fields appearing in the Lagrangian of the theory. However, often the bare monopole operator has to be supplemented by a number of fundamental fields in order to construct a gauge invariant operator. \\

Our aim is to find the conformal dimensions of such monopole operators in the ABJM model. In particular we are interested in small Chern-Simons level $k$, as for $k=1,2$ we expect supersymmetry enhancement in this theory. For small $k$, however, ABJM is strongly coupled and thus we cannot employ perturbation theory since there is no small parameter. The key idea to circumvent this obstacle is to add a Yang-Mills term for the gauge fields in the action which introduces another coupling $g$ as a second parameter besides $k$. This also requires adding dynamical fields in the adjoint representation in order to preserve $\superN=3$ supersymmetry and an $\grSU(2)_R$ subgroup of the $\grSU(4)_R$ group of ABJM (which has $\superN=6$ supersymmetry). \\

The coupling $g$ is dimensionful and not a parameter we can dial. Since the Yang-Mills term is irrelevant in three dimensions, there is a renormalization group (RG) flow from the ultraviolet (UV), where the theory is free (vanishing $g$) to a conformal infrared (IR) fixed point (divergent $g$). This RG flow appears naturally in the brane construction of the ABJM model and in fact is essential for understanding how a pure Chern-Simons theory such as ABJM can arise from D-branes that support Yang-Mills theories. \\

In the IR, the Yang-Mills terms as well as the kinetic terms for the adjoint matter fields drop out of the action and the equations of motion for the latter degenerate into constraint equations, allowing us to integrate them out. In this way one recovers the ABJM theory. \\

Now the idea is to perform all relevant computations in the far UV where $g$ is small and the theory weakly coupled, then flow to the IR. The scaling dimensions of monopole operators are of course only well-defined at the IR fixed point where the theory is conformal, so they cannot be determined directly in this fashion.  However, we can instead compute a quantity that is preserved along the RG flow and related by supersymmetry to the scaling dimensions in the IR: the non-abelian R-charges of the monopole operators. Their one-loop value is exact and preserved along the RG flow because non-abelian representations cannot change continuously and therefore cannot depend (non-trivially) on $g$. \\

Let us describe the computation of the non-abelian R-charges. In the UV there is a separation of scales between the BPS background that inserts a flux at a spacetime point in accordance with \eqref{eqn:Dirac} -- recall that its magnitude is constrained by the Dirac quantization condition -- and the typical size of quantum fluctuations of fields. Therefore we can treat the monopole operator as a classical background. For this background to satisfy the BPS condition the scalar fields $\phi_i$ and $\hat{\phi}_i$, which are in the same $\superN=3$ vector multiplets as the gauge fields $A_\mu$ and $\hat{A}_\mu$, respectively, need to be turned on. A possible choice of scalar background in radial quantization on $\Reals\times\Sphere^2$ is
\be \label{eqn:bps-background-static}
  \phi_i = - \hat{\phi}_i = - \frac{H}{2} \delta_{i3} \; .
\ee
These scalar fields transform in the $\rep{3}$ of $\grSU(2)_R$ and therefore a non-zero expectation value breaks $\grSU(2)_R$ to $\grU(1)_R$. \\

As a first step we will find the one-loop $\grU(1)_R$ charge of the monopole operator described by this fixed background. This is done by computing the normal ordering constant for the $\grU(1)_R$ charge operator. Such a calculation has been performed in a different context in \cite{Borokhov:2002ib,Borokhov:2002cg}. We will present the argument as applicable to our case in \secref{sec:U1}. The result will turn out to be expression \eqref{eqn:induced} above or more concretely \eqref{eqn:U1R-ABJM-result} for $\superN=3$ $\grU(N)\times\grU(N)$ gauge theory with hyper multiplets in the bifundamental. \\

These two formulas are related as follows. From \tabref{tab:R-charges}, or equivalently from the explicit expression for the R-current \eqref{eqn:U1R-Noether-current-fermions}, one can read off the R-charges $y(\zeta^A) = y(\omega_A) = -\half$ ($A=1,\ldots,N_f$) for the hyper multiplet fermions and $y(\chisigma) = y(\hchisigma) = 1$, $y(\chiphi) = y(\hchiphi) = 0$ for those in the vector multiplets. In the expression for the $\grU(1)_R$ charge of the BPS monopole \eqref{eqn:induced}, these R-charges are weighted by $\sum_{r,s} \abs{q_r - q_s}$ and the result is given by
\be
  Q_R^{\mathrm{mon}} = \Half \Bigbrk{ 2 N_f \cdot \bigabs{-\half}  - 2 \cdot \abs{+ 1} } \sum_{r,s} \abs{q_r-q_s}  = \lrbrk{\frac{N_f}{2}-1} \sum_{r,s} \abs{q_r-q_s} \; ,
\ee
as found in \eqref{eqn:U1R-ABJM-result}. For the ABJM model we have $N_f = 2$ and hence $Q_R^{\mathrm{mon}}=0$. \\

These $\grU(1)_R$ charges might in principle be renormalized as we flow to the IR, since abelian charges can vary continuously. To exclude this possibility, we need to find the non-abelian $\grSU(2)_R$ charge of the monopole operator by taking into account the bosonic zero modes of the background \eqref{eqn:bps-background-static}. These zero modes are described by a unit vector $\vec{n}$ on the two-sphere $\grSU(2)_R/\grU(1)_R$. In other words, we will treat the $\grSU(2)_R$ orientation $\vec{n}$ of the scalar field as a collective coordinate of the BPS background and quantize its motion. To this end we consider a more general background than \eqref{eqn:bps-background-static}, which is allowed to depend on (Euclidean) time $\tau$
\be \label{eqn:bps-background-rotating}
  \phi_i = - \hat{\phi}_i = - \frac{H}{2} n_i(\tau) \; .
\ee

The rotation of the background is assumed to be adiabatic such that it cannot excite finite energy quantum fluctuations around the background. In \secref{sec:SU2} we will compute the quantum mechanical effective action for the collective coordinate $\vec{n}(\tau)$ and find the allowed $\grSU(2)_R$ representations. These representations are precisely the $\grSU(2)_R$ spectrum of BPS monopole operators. \\

If there were no interactions between the bosonic zero modes and other fields in the Lagrangian, the collective coordinate would simply be described by a free particle on a sphere. Such a particle -- and therefore the monopole operator -- could be in any representation of $\grSU(2)_R$. The crucial point is, however, that the collective coordinate $\vec{n}$ is not free, but subject to interactions with the fermions of the theory. It turns out that the induced interaction term is a coupling of $\vec{n}$ to a Dirac monopole of magnetic charge $h\in\Integers$ on the collective coordinate moduli space (not to be confused with the spacetime monopole background). The charge $h$ depends on the background flux $H$ and the field content of the theory. In \secref{sec:SU2} we derive
\be
  h \eq (N_f - 2) \, q_{\mathrm{tot}} = (N_f - 2) \sum_{r,s} \abs{q_r - q_s} \; , \\[-7mm] \nn
\ee
see \eqref{eqn:induced-mag-charge}, which is nothing but $h = 2 Q_R^{\mathrm{mon}}$. \\

The equation of motion for the collective coordinates $\vec{n}$ has the structure
\be \label{eqn:eom-coll-coord}
  M \, \ddot{\vec{n}} + M (\dot{\vec{n}})^2 \, \vec{n} + \frac{h}{2} \, \vec{n} \times \dot{\vec{n}} = 0 \; ,
\ee
which allows for a simple particle interpretation. The first term is the usual kinetic term while the second term represents a constraining force that keeps the particle on the sphere $\vec{n}^2 = 1$. If we assign unit electric charge to the particle, then the third term is the Lorentz force $\dot{\vec{n}}\times \vec{B}$ due to a monopole field $\vec{B} = \frac{h}{2} \vec{n}$. The $\grSU(2)_R$ charge is given by the conserved angular momentum
\be
  \vec{L} = M \, \vec{n}\times \dot{\vec{n}} - \frac{h}{2} \, \vec{n} \; .
\ee
The presence of the second term forces the quantized angular momentum to have an orbital quantum number of at least $l=\frac{\abs{h}}{2}$, the possible representations being $\vec{L}^2 = l(l+1)$ with $l=\frac{\abs{h}}{2}, \frac{\abs{h}}{2}+1, \frac{\abs{h}}{2}+2, \ldots$. Hence, a non-zero coupling $h$ will necessarily give the monopole a non-trivial $\grSU(2)_R$ charge. The ABJM theory is special (as compared to theories with different field content) in that all contributions to the effective $h$ cancel each other, and thus it allows $\grSU(2)_R$ singlet monopoles which do not contribute to the IR dimensions of gauge invariant operators.

%%%%%%%%%%%%%%%%%%%%%%%%%%%%%%%%%%%%%%%%%%%%%%%%%%%%%%%%%%%%%%%%%%%%%%%%%%%
%%%%%%%%%%%%%%%%%%%%%%%%%%%%%%%%%%%%%%%%%%%%%%%%%%%%%%%%%%%%%%%%%%%%%%%%%%%
\section{$\superN=3$ Chern-Simons Yang-Mills theory}
\label{sec:CSYM}

We will study $\superN=3$ supersymmetric $\grU(N)\times\hat{\grU}(N)$ 3-d gauge theory coupled to bifundamental matter fields and $\grSU(N_f)_{\mathrm{fl}}$ flavor symmetry. For $N_f=1$ and $N_f=2$ this model is the UV completion of the $\superN=4$ model of GW \cite{Gaiotto:2008sd} and the $\superN=6$ model of ABJM \cite{Aharony:2008ug}, respectively. \\

%%%%%%%%%%%%%%%%%%%%%%%%%%%%%%%%%%%%%%%%%%%%%%%%%%%%%%%%%%%%%%%%%%%%%%%%%%%
\subsection{Action and supersymmetry transformations}

The complete field content is listed in \tabref{tab:fields}. We use $\superN=2$ superfields and our notation is explained in \appref{sec:notation}. In the gauge sector we have two vector superfields $\VV = (\sigma, A_\mu, \chi_\sigma, \auxD)$ and $\hat{\VV} = (\hat{\sigma}, \hat{A}_\mu, \hat{\chi}_\sigma, \hat{\auxD})$ and two chiral superfields $\Phi = (\phi, \chi_\phi, F_\phi)$ and $\hat{\Phi} = (\hat{\phi}, \hat{\chi}_\phi, \hat{F}_\phi)$ in the adjoint of the two gauge groups $\grU(N)$ and $\hat{\grU}(N)$, respectively, which together comprise two $\superN=3$ gauge multiplets. In the matter sector we have $N_f$ chiral superfields $\ZZ^A = (Z^A, \zeta^A, F^A)$ and $\WW_A = (W_A, \omega_A, G_A)$ in the gauge representations $(\rep{N},\rep{\bar{N}})$ and $(\rep{\bar{N}},\rep{N})$, respectively. \\

\begin{table}
\begin{center}
\renewcommand{\arraystretch}{1.4}
\begin{tabular}{|l|c|c:c:c:c|c:c:c:c|c|} \hline
\multirow{2}{*}{\parbox{20mm}{manifest\\symmetry}} & \multirow{2}{*}{\parbox{11mm}{Super\\fields}} & \multicolumn{9}{|c|}{Component fields} \\ \cline{3-11}
 & & \multicolumn{4}{|c|}{dynamical in the IR} & \multicolumn{4}{|c|}{auxiliary in the IR} & aux. \\ \hline
\multirow{10}{*}{$\grU(1)_R\times\grSU(N_f)_{\mathrm{fl}}$} & $\VV$ & \multicolumn{4}{|c|}{$A$} & $\sigma$ & & $\chisigma$, $\chi_\sigma^\dagger$ & & $\auxD$ \\
& $\hat{\VV}$ & \multicolumn{4}{|c|}{$\hat{A}$} & & $\hat{\sigma}$ & & $\hchisigma$, $\hat{\chi}_\sigma^\dagger$ & $\hat{\auxD}$ \\ \cline{2-11}
& $\Phi$ & \multicolumn{4}{|c|}{} & $\phi$ & & $\chiphi$ & & $F_\phi$ \\
& $\bar{\Phi}$ & \multicolumn{4}{|c|}{} & $\phi^\dagger$ & & $\chi_\phi^\dagger$ & & $F_\phi^\dagger$ \\
& $\hat{\Phi}$ & \multicolumn{4}{|c|}{} & & $\hat{\phi}$ & & $\hchiphi$ & $\hat{F}_\phi$ \\
& $\hat{\bar{\Phi}}$ & \multicolumn{4}{|c|}{} & & $\hat{\phi}^\dagger$ & & $\hat{\chi}_\phi^\dagger$ & $\hat{F}_\phi^\dagger$ \\ \cline{2-11}
& $\ZZ^A$ & $Z^A$ & & $\zeta^A$ & & & & & & $F^A$ \\
& $\bar{\ZZ}_A$ & & $Z^\dagger_A$ & & $\zeta^\dagger_A$ & & & & & $F^\dagger_A$ \\
& $\WW_A$ & & $W_A$ & & $\omega_A$ & & & & & $G_A$ \\
& $\bar{\WW}^A$ & $W^{\dagger A}$ & & $\omega^{\dagger A}$ & & & & & & $G^{\dagger A}$ \\ \hline
$\grSU(2)_R\times\grSU(N_f)_{\mathrm{fl}}$ & & $X^{Aa}$ & $X^\dagger_{Aa}$ & $\xi^{Aa}$ & $\xi^\dagger_{Aa}$ & $\phi_i$ & $\hat{\phi}_i$ & $\lambda^{ab}$ & $\hat{\lambda}^{ab}$ \\ \cline{1-10}
\end{tabular}
\renewcommand{\arraystretch}{1}
\end{center}
\caption{\textbf{\mathversion{bold} Field content of $\superN=3$ Chern-Simons Yang-Mills theory.} Each row (except the last one) shows the components of one $\superN=2$ superfield ordered into columns according to whether they are dynamical or auxiliary. Within each of these columns they have been arranged such that one can read off from the last row which components join to form $\grSU(2)_R$ multiplets. $A=1,\ldots,N_f$ is a $\grSU(N_f)$ flavor index, $a=1,2$ is a $\grSU(2)_R$ spinor index, and $i=1,2,3$ is a $\grSU(2)_R$ vector index.}
\label{tab:fields}
\end{table}

We write down the $\superN=3$ action on $\Reals^{1,2}$ with signature $(-,+,+)$. It consists of five parts:
\be
  \Action = \Action_{\mathrm{CS}} + \Action_{\mathrm{YM}} + \Action_{\mathrm{adj}} + \Action_{\mathrm{mat}} + \Action_{\mathrm{pot}} \; .
\ee
First of all there are Chern-Simons terms
\be
  \Action_{\mathrm{CS}} \eq -i \frac{k}{8\pi} \int\!d^3x\,d^4\theta \int_0^1 ds\: \tr \Bigsbrk{
      \VV \bar{D}^\alpha \Bigbrk{ e^{s \VV} D_\alpha e^{-s \VV} } -
      \hat{\VV} \bar{D}^\alpha \Bigbrk{ e^{s \hat{\VV}} D_\alpha e^{-s \hat{\VV}} }
      } \; ,
\ee
with opposite levels $k$ and $-k$ for the two gauge group factors. The $s$ integral is nothing but a convenient way of writing the non-abelian Chern-Simons action. Secondly, there is a Yang-Mills term
\be
  \Action_{\mathrm{YM}} \eq \frac{1}{4g^2} \int\!d^3x\,d^2\theta\: \tr \Bigsbrk{ \mathcal{U}^\alpha \mathcal{U}_\alpha + \hat{\mathcal{U}}^\alpha \hat{\mathcal{U}}_\alpha } \; ,
\ee
which introduces a coupling $g$ of mass dimension $\half$. The super field strength is given by $\mathcal{U}_\alpha = \Quarter \bar{D}^2 e^{\VV} D_\alpha e^{-\VV}$ and similarly for $\hat{\mathcal{U}}$. The last term in the gauge sector is given by the kinetic terms for the adjoint scalar fields
\be
  \Action_{\mathrm{adj}} = \frac{1}{g^2} \int d^3x\,d^4\theta\: \tr \Bigsbrk{
      - \bar{\Phi} e^{-\VV} \Phi e^{\VV}
      - \hat{\bar{\Phi}} e^{-\hat{\VV}} \hat{\Phi} e^{\hat{\VV}}
   } \; .
\ee
In the matter sector we have the minimally coupled action for the bifundamental fields
\be
  \Action_{\mathrm{mat}} = \int d^3x\,d^4\theta\: \tr \Bigsbrk{ - \bar{\ZZ}_A e^{-\VV} \ZZ^A e^{\hat{\VV}} - \bar{\WW}^A e^{-\hat{\VV}} \WW_A e^{\VV} }
\ee
and a super potential term
\be
  \Action_{\mathrm{pot}} = \int\!d^3xd^2\theta\: \superW - \int\!d^3xd^2\bar{\theta}\: \bar{\superW}
\ee
with
\be
  \superW \eq \tr \bigbrk{ \Phi \ZZ^A \WW_A + \hat{\Phi} \WW_A \ZZ^A }
            + \frac{k}{8\pi} \tr \bigbrk{ \Phi \Phi - \hat{\Phi} \hat{\Phi} } \; , \\
  \bar{\superW} \eq \tr \bigbrk{ \bar{\Phi} \bar{\WW}^A \bar{\ZZ}_A + \hat{\bar{\Phi}} \bar{\ZZ}_A \bar{\WW}^A }
            + \frac{k}{8\pi} \tr \bigbrk{ \bar{\Phi} \bar{\Phi} - \hat{\bar{\Phi}} \hat{\bar{\Phi}} } \; .
\ee
This theory is not conformal and will flow to an IR fixed point which is strongly coupled unless $k$ is large. At the fixed point $g$ diverges which renders the gauge fields and the adjoint scalars non-dynamical. If we integrate out $\Phi$ and $\hat{\Phi}$, we recover the ABJM superpotential for $N_f=2$. This theory has enhanced flavor symmetry, $\grSU(2)_{\mathrm{fl}}\times\grSU(2)_{\mathrm{fl}}$, under which $\ZZ^A$ and $\WW_A$ transform separately. \\

For our computation below we need the action in terms of component fields. To this end we perform the Grassmann integrals in the action and integrate out the auxiliary fields $\auxD$, $\hat{\auxD}$, $F^A$, $G_A$, $F_\phi$ and $\hat{F}_\phi$. The remaining component fields can be arranged into $\grSU(2)_R$ multiplets as follows.

The adjoint matter fields constitute two scalars\footnote{The lower/upper index is the row/column index.}
\be
  \phi^a_b = \phi_i (\sigma_i)^a_b
  = \matr{cc}{
     -\sigma & \phi^\dagger \\
     \phi & \sigma
  }
  \comma
  \hat{\phi}^a_b = \hat{\phi}_i (\sigma_i)^a_b
   = \matr{cc}{
     \hat{\sigma} & \hat{\phi}^\dagger \\
     \hat{\phi} & -\hat{\sigma}
  } \; ,
\ee
%\be
%  \phi^a_b = \phi_i (\sigma_i)^a_b
%  \comma
%  \phi_i = \matr{c}{
%     \half \bigbrk{\phi^\dagger + \phi} \\[1mm]
%     \ihalf \bigbrk{\phi^\dagger - \phi} \\[1mm]
%     - \sigma
%  }
%  \comma
%  \hat{\phi}^a_b = \hat{\phi}_i (\sigma_i)^a_b
%  \comma
%  \hat{\phi}_i = \matr{c}{
%     \half \bigbrk{\hat{\phi}^\dagger + \hat{\phi}} \\[1mm]
%     \ihalf \bigbrk{\hat{\phi}^\dagger - \hat{\phi}} \\[1mm]
%     \hat{\sigma}
%  }
%  \; ,
%\ee
transforming in the $\rep{3}$ of $\grSU(2)_R$, and two fermions
\be
    \lambda^{ab} = \matr{cc}{
    \chisigma \, e^{-i\pi/4} & \chi_\phi^\dagger \, e^{-i\pi/4} \\
    \chiphi \, e^{+i\pi/4} & -\chi_\sigma^\dagger \, e^{+i\pi/4}
  }
  \comma
  \hat{\lambda}^{ab} = \matr{cc}{
    \hchisigma \, e^{-i\pi/4} & -\hat{\chi}_\phi^\dagger \, e^{-i\pi/4} \\
    -\hchiphi \, e^{+i\pi/4} & -\hat{\chi}_\sigma^\dagger \, e^{+i\pi/4}
  }
  \; ,
\ee
transforming in the reducible representation $\rep{2}\times\rep{2} = \rep{3}+\rep{1}$ of $\grSU(2)_R$, which implies that $\lambda^{ab}$ is neither symmetric nor anti-symmetric in its indices. These fields satisfy
\be \label{eqn:relations}
  (\lambda^{ab})^* = -\lambda_{ab} = -\levi_{ac} \levi_{bd} \lambda^{cd}
  \comma
  (\phi^a_b)^* = \phi^b_a = \levi_{ac} \levi^{bd} \phi^c_d
\ee
and the same for the hatted fields.

The bifundamental matter fields can be grouped into $N_f$ doublets of the $\grSU(2)_R$ symmetry group in the following way:
\be \label{eqn:X-SU2R}
  X^{Aa} = \matr{c}{Z^A \\ W^{\dagger A}}
  \comma
  X^\dagger_{Aa} = \matr{c}{Z^\dagger_A \\ W_A}
  \; ,
\ee
and
\be \label{eqn:xi-SU2R}
  \xi^{Aa} = \matr{c}{ \omega^{\dagger A} \, e^{i\pi/4} \\ \zeta^A \, e^{-i\pi/4} }
  \comma
  \xi^\dagger_{Aa} = \matr{c}{ \omega_A \, e^{-i\pi/4} \\ \zeta^\dagger_A \, e^{i\pi/4} }
  \; .
\ee

The component action with manifest $\grSU(2)_R$ symmetry and the corresponding supersymmetry transformations are given in \appref{sec:N3-YM-CS-R12}. One observes that in the IR, where $g$ becomes large, the Yang-Mills terms disappears together with the kinetic terms for $\phi$, $\hat{\phi}$, $\lambda$ and $\hat{\lambda}$, and we can integrate out these fields. Doing this for $N_f=2$ we end up with ABJM theory. The ABJM Lagrangian with manifest $\grSU(4)_R$ symmetry \cite{Benna:2008zy} is recovered when one defines $Y^A = \{ X^{11}, X^{21}, X^{12}, X^{22} \}$ and $\psi_A = \{ -\xi^{22}, \xi^{12}, \xi^{21}, -\xi^{11} \}$. \\

As outlined in the overview, \secref{sec:heart}, we will carry out our computations in the far UV region of the radially quantized theory. The point of going to the far UV is that the theory becomes perturbative in $g$ and we can find the quantum R-charges from a one-loop computation. This one-loop result is in fact the exact answer because $\grSU(2)_R$ is preserved along the flow from small to large $g$ and non-abelian representations cannot change continuously. We will use radial quantization because we are interested in the spectrum of conformal dimensions in the IR theory, which is related to the energy spectrum by the operator state correspondence.

The steps required for deriving the action relevant for radial quantization are as follows. First we perform a Wick rotation from $\Reals^{1,2}$ to $\Reals^{3}$ by defining Euclidean coordinates $(x^1,x^2,x^3) = (x^1,x^2,i x^0)$, then we change to polar coordinates $(r,\theta,\varphi)$ and finally we introduce a new radial variable $\tau$ by setting $r = e^\tau$. The result is a theory on $\Reals\times\Sphere^2$ described by the coordinates $(\tau,\theta,\varphi)$, where $\tau\in\Reals$ is the ``Euclidean time''. The change of the coordinates is accompanied by a Weyl rescaling of the fields according to
\be
 \mathcal{A} = e^{-\dim(\mathcal{A})\tau} \tilde{\mathcal{A}} \; ,
\ee
where $\mathcal{A}$ is a generic field of dimension $\dim(\mathcal{A})$ on $\Reals^3$ and $\tilde{\mathcal{A}}$ is the field we use on $\Reals\times\Sphere^2$. After the transformation we will drop the tildes in order to avoid cluttered notation. Since the theory is not conformal, the action will change under these rescaling. In addition to the mass terms which are generated even in the conformal case, every factor of the coupling $g$ will turn into
\be \label{eqn:gtilde-euclid}
  \tg = e^{\tau/2} \, g \; .
\ee
This relation makes the RG-flow explicit and relates it to the ``time'' on $\Reals\times\Sphere^2$. In the infinite past the effective Yang-Mills coupling $\tg$ vanishes, and it grows without bound toward future infinity. To compute in the UV therefore means to work at $\tau\to-\infty$ and to flow to the IR means to send $\tau\to+\infty$.

Now we are ready to give the complete action of $\superN=3$ Chern-Simons Yang-Mills theory on $\Reals\times\Sphere^2$. In addition to the manipulations just described we have rescaled $\lambda \to g \lambda$ and $\hat{\lambda} \to g \hat{\lambda}$ (before the Weyl rescaling) which is the appropriate scaling for fluctuations in the UV. We do \emph{not} perform a similar rescaling of the other fields in the gauge sector, $A$, $\hat{A}$, $\phi$, or $\hat{\phi}$, since these fields will later provide a large classical background. Introducing finally the rescaled Chern-Simons coupling $\kappa = \frac{k}{4\pi}$, the kinetic part of the action reads\footnote{Suppressed indices are assumed to be in the standard positions as defined in \eqref{eqn:X-SU2R} and \eqref{eqn:xi-SU2R}. The indices of the Pauli matrices are placed accordingly, e.g. $X \sigma_i X^\dagger \equiv X^{Aa} (\sigma_i)_a{}^b X^\dagger_{Ab}$ or $X^\dagger \sigma_i X \equiv X^\dagger_{Aa} (\sigma_i)^a{}_b X^{Ab}$. And by definition we have $(\sigma_i)_a{}^b = \sigma_i$ and $(\sigma_i)^a{}_b = \sigma_i^\trans$.}
\be \label{eqn:action-CS-SYM-spinor-euclidean-kin}
  \Action_{\mathrm{kin}}^E \eq \int\!d\tau\,d\Omega\: \tr \Bigsbrk{
   + \tfrac{1}{2\tg^2} F^{mn} F_{mn}
   - \kappa \, i \levi^{mnk} \bigbrk{
            A_m \partial_n A_k
            + \tfrac{2i}{3} A_m A_n A_k }
   \nl\hspace{21mm}
   + \tfrac{1}{2\tg^2} \hat{F}^{mn} \hat{F}_{mn}
   + \kappa \, i \levi^{mnk} \bigbrk{
            \hat{A}_m \partial_n \hat{A}_k
            + \tfrac{2i}{3} \hat{A}_m \hat{A}_n \hat{A}_k }
   \nl[1mm]\hspace{21mm}
   + \deriD_m X^\dagger \deriD^m X
   + \quarter X^\dagger X
   - i \xi^\dagger \slashed{\deriD} \xi
   \nl[1mm]\hspace{21mm}
   + \tfrac{1}{2\tg^2} \deriD_m \phi^a_b \deriD^m \phi^b_a
   + \tfrac{\kappa^2\tg^2}{2} \, \phi^a_b \phi^b_a
   + \tfrac{1}{2\tg^2} \deriD_m \hat{\phi}^a_b \deriD^m \hat{\phi}^b_a
   + \tfrac{\kappa^2\tg^2}{2} \, \hat{\phi}^a_b \hat{\phi}^b_a
   \nl[1mm]\hspace{21mm}
   + \ihalf \lambda^{ab} \slashed{\deriD} \lambda_{ab}
   + \tfrac{\kappa \tg^2}{2} \, i \lambda^{ab} \lambda_{ba}
   + \ihalf \hat{\lambda}^{ab} \slashed{\deriD} \hat{\lambda}_{ab}
   - \tfrac{\kappa \tg^2}{2} \, i \hat{\lambda}^{ab} \hat{\lambda}_{ba}
  }
\ee
and the interaction terms are given by
\be \label{eqn:action-CS-SYM-spinor-euclidean-int}
  \Action_{\mathrm{int}}^E \eq \int\!d\tau\,d\Omega\: \tr \Bigsbrk{
   + \kappa \tg^2 \, X^\dagger_a \phi^a_b X^b
   - \kappa \tg^2 \, X^a \hat{\phi}^b_a X^\dagger_b
   + i \xi^\dagger_a \phi^a_b \xi^b
   + i \xi^a \hat{\phi}^b_a \xi^\dagger_b
   \nl[-2mm]\hspace{21mm}
   - \tg \, \levi_{ac} \lambda^{cb} X^a \xi^\dagger_b
   + \tg \, \levi^{ac} \lambda_{cb} \xi^b X^\dagger_a
   + \tg \, \levi_{ac} \hat{\lambda}^{cb} \xi^\dagger_b X^a
   - \tg \, \levi^{ac} \hat{\lambda}_{cb} X^\dagger_a \xi^b
   \nl[1mm]\hspace{21mm}
   - \tfrac{\kappa}{6} \phi^a_b \comm{\phi^b_c}{\phi^c_a}
   - \tfrac{\kappa}{6} \hat{\phi}^a_b \comm{\hat{\phi}^b_c}{\hat{\phi}^c_a}
   + \ihalf \lambda_{ab} \comm{\phi^b_c}{\lambda^{ac}}
   - \ihalf \hat{\lambda}_{ab} \comm{\hat{\phi}^b_c}{\hat{\lambda}^{ac}}
   \nl[1mm]\hspace{21mm}
   + \tfrac{\tg^2}{4} (X \sigma_i X^\dagger) (X \sigma_i X^\dagger)
   + \tfrac{\tg^2}{4} (X^\dagger \sigma_i X) (X^\dagger \sigma_i X)
  \nl[1mm]\hspace{21mm}
   + \tfrac{1}{2} (X X^\dagger) \phi^a_b \phi^b_a
   + \tfrac{1}{2} (X^\dagger X) \hat{\phi}^a_b \hat{\phi}^b_a
   + X^\dagger_{Aa} \phi^b_c X^{Aa} \hat{\phi}^c_b
   \nl\hspace{21mm}
   - \tfrac{1}{8\tg^2} \comm{\phi^a_b}{\phi^c_d} \comm{\phi^b_a}{\phi^d_c}
   - \tfrac{1}{8\tg^2} \comm{\hat{\phi}^a_b}{\hat{\phi}^c_d} \comm{\hat{\phi}^b_a}{\hat{\phi}^d_c}
  } \; .
\ee
The covariant derivatives are given by $\deriD_m X = \nabla_m X + i A_m X - i X \hat{A}_m$ etc. We also translate the supersymmetry variations from flat Lorentzian space as given in \appref{sec:N3-YM-CS-R12} to Euclidean $\Reals\times\Sphere^2$. They are conveniently expressed in terms of a rescaled parameter
\be
  \tilde{\eps}_{ab}(\tau) = \eps_{ab} e^{-\tau/2} \; .
\ee
In the following we will use the $\tau$ dependent parameter; however, we will drop the tilde for notational simplicity. This parameter satisfies the Killing spinor equation
\be \label{eqn:Killing}
  \nabla_m \eps = -\half \gamma_m \gamma^\tau \eps \; ,
\ee
which is the curved spacetime generalization of the usual condition of (covariant) constancy that the supersymmetry variation parameter obeys in flat space.
The $\superN = 3$ supersymmetry transformations read
\be \label{eqn:SusyLambda}
  \delta A_m          \eq - \tfrac{i\tg}{2} \eps_{ab} \gamma_m \lambda^{ab} \; , \\
  \delta \lambda^{ab} \eq   \tfrac{i}{2\tg} \levi^{mnk} F_{mn} \gamma_k \eps^{ab}
                          - \tfrac{i}{\tg} \slashed{\deriD} \phi^b_c \eps^{ac}
                          - \tfrac{2i}{3\tg} \phi^b_c \slashed{\nabla} \eps^{ac}
                          + \tfrac{i}{2\tg} \comm{\phi^b_c}{\phi^c_d} \eps^{ad}
                          + \kappa \tg \, i \phi^b_c \eps^{ac} \nl
                          + \tg \, i X^a X^\dagger_c \eps^{cb}
                          - \tfrac{i\tg}{2} (X X^\dagger) \eps^{ab} \; , \nn \\
  \delta \phi^a_b     \eq - \tg \eps_{cb} \lambda^{ca}
                          + \tfrac{\tg}{2} \delta^a_b \eps_{cd} \lambda^{cd} \; , \nn
\ee
\be \label{eqn:SusyLambdaHat}
  \delta \hat{A}_m          \eq - \tfrac{i\tg}{2} \eps_{ab} \gamma_m \hat{\lambda}^{ab} \; , \\
  \delta \hat{\lambda}^{ab} \eq   \tfrac{i}{2\tg} \levi^{mnk} \hat{F}_{mn} \gamma_k \eps^{ab}
                                + \tfrac{i}{\tg} \slashed{\deriD} \hat{\phi}^b_c \eps^{ac}
                                + \tfrac{2i}{3\tg} \hat{\phi}^b_c \slashed{\nabla} \eps^{ac}
                                + \tfrac{i}{2\tg} \comm{\hat{\phi}^b_c}{\hat{\phi}^c_d} \eps^{ad}
                                + \kappa \tg \, i \hat{\phi}^b_c \eps^{ac} \nl
                                - \tg \, i \eps^{bc} X^\dagger_c X^a
                                + \tfrac{i\tg}{2} (X^\dagger X) \eps^{ab} \; , \nn \\
  \delta \hat{\phi}^a_b     \eq - \tg \eps_{cb} \hat{\lambda}^{ca}
                                + \tfrac{\tg}{2} \delta^a_b \eps_{cd} \hat{\lambda}^{cd} \; , \nn
\ee
\begin{align}  \label{eqn:SusyXi}
  \delta X^{Aa}           & = - i \eps^a_b \xi^{Ab} \; , &
  \delta \xi^{Aa}         & = \slashed{\deriD} X^{Ab} \eps^a_b
                              + \tfrac{1}{3} X^{Ab} \slashed{\nabla} \eps^a_b
                              + \phi^a_b \eps^b_c X^{Ac}
                              + X^{Ac} \eps^b_c \hat{\phi}^a_b \; , \\
  \delta X^\dagger_{Aa}   & = - i \xi^\dagger_{Ab} \eps^b_a \; , &
  \delta \xi^\dagger_{Aa} & = \slashed{\deriD} X^\dagger_{Ab} \eps^b_a
                              + \tfrac{1}{3} X^\dagger_{Ab} \slashed{\nabla} \eps^b_a
                              + \hat{\phi}^b_a \eps^c_b X^\dagger_{Ac}
                              + X^\dagger_{Ac} \eps^c_b \phi^b_a \; . \nn
\end{align}

\paragraph{$\grSU(2)_R$ and $\grU(1)_R$ charge.} Fundamental and anti-fundamental $\grSU(2)_R$ indices transform under infinitesimal rotations as
\be \label{eqn:SU2R-trafo}
  \delta \mathcal{A}^a = i \eps^a_b \mathcal{A}^b
  \comma
  \delta \mathcal{A}_a = - i \eps^b_a \mathcal{A}^\dagger_b
  \; ,
\ee
where $\mathcal{A}$ represents a generic field. The Noether current is
\be \label{eqn:SU2R-Noether-current}
  J^{\mu b}_a &\sim& \tr \Bigsbrk{
                               i X^\dagger_{Aa} \lrderiD^\mu X^{Ab}
                             + i \phi_a^c \lrderiD^\mu \phi_c^b
                             + i \hat{\phi}_a^c \lrderiD^\mu \hat{\phi}_c^b
                             - \xi^\dagger_{Aa} \gamma^\mu \xi^{Ab} \nl\hspace{5mm}
                             + \half \lambda_{ac} \gamma^\mu \lambda^{bc}
                             + \half \lambda_{ca} \gamma^\mu \lambda^{cb}
                             + \half \hat{\lambda}_{ac} \gamma^\mu \hat{\lambda}^{bc}
                             + \half \hat{\lambda}_{ca} \gamma^\mu \hat{\lambda}^{cb}
                           }
  \; .
\ee
In \secref{sec:U1} we will be dealing with the $\grU(1)_R$ component of this current which is related to the transformation \eqref{eqn:SU2R-trafo} with $\eps^a_b \sim (\sigma_3)^a_b$. Hence the current is the contraction of \eqref{eqn:SU2R-Noether-current} with $(\sigma_3)^a_b$.
The part due to the fermions is given by
\be \label{eqn:U1R-Noether-current-fermions}
  J^\mu &\sim& \tr \Bigsbrk{ - \half \, \zeta^\dagger_A \gamma^\mu \zeta^A
                             - \half \, \omega^{\dagger A} \gamma^\mu \omega_A
                             + \chi_\sigma^\dagger \gamma^\mu \chi_\sigma
                             + \hat{\chi}_\sigma^\dagger \gamma^\mu \hat{\chi}_\sigma
                           }
  \; ,
\ee
where we have reverted back to the $\grU(1)_R\times\grSU(N_f)_{\mathrm{fl}}$ fields, see \tabref{tab:fields}. From this expression we read off the $\grU(1)_R$ charges of the fermions as given in \tabref{tab:R-charges}.

\begin{table}
\begin{center}
\renewcommand{\arraystretch}{1.4}
\begin{tabular}{c|c|c|c|c|c}
$y_i$ & $-1$ & $-\half$ & $0$ & $+\half$ & $+1$ \\ \hline
field &
$\chi_\sigma^\dagger$,
$\hat{\chi}_\sigma^\dagger$
&
$\zeta^A$,
$\omega_A$,
&
$\chiphi$,
$\hchiphi$,
$\chi_\phi^\dagger$,
$\hat{\chi}_\phi^\dagger$
&
$\zeta^\dagger_A$,
$\omega^{\dagger A}$
&
$\chisigma$,
$\hchisigma$
\end{tabular}
\renewcommand{\arraystretch}{1}
\end{center}
\caption{\textbf{\mathversion{bold} $\grU(1)_R$ charges of the fermion fields.} These numbers show what the sum (chiral only) of R-charges vanishes precisely for two flavors, $A=1,2$.}
\label{tab:R-charges}
\end{table}

%%%%%%%%%%%%%%%%%%%%%%%%%%%%%%%%%%%%%%%%%%%%%%%%%%%%%%%%%%%%%%%%%%%%%%%%%%%
\subsection{Classical monopole solution}

Since we are interested in BPS monopoles, our first task is to find a classical BPS solution with flux emanating from a point in spacetime. Starting from the gauge field configuration of a Dirac monopole in $\Reals^3$ given in \eqref{eqn:Dirac} and performing the Weyl rescaling appropriate for fields on $\Reals\times\Sphere^2$  as described above, we find that the dependence on the radial coordinate disappears from the gauge potential. In fact the Hodge dual of the corresponding field strength,  $\levi^{mnk} F_{mn}$, is constant in magnitude and purely radial (i.e. only its $\tau$ component is non-vanishing).

In order to show that there is indeed a BPS solution with such a Dirac monopole potential,  let us examine in detail the supersymmetry variations of $\lambda^{ab} $. It contains terms of order $\tg$ and ones of order $\tg^{-1}$, but since we are looking for a solution that is supersymmetric along the whole RG flow they should cancel separately. We will also assume that all background fields (in our choice of gauge) are valued in the Cartan subalgebras of the gauge group factors, so that all commutators vanish.

Focusing on the terms of order $\tg^{-1}$ for now, we thus want the sum of the first three terms in  $\delta \lambda^{ab}$, as given in \eqref{eqn:SusyLambda}, to vanish for an appropriate, non-trivial choice of supersymmetry variation parameter. The fact that  $\levi^{mnk} F_{mn}$ is constant suggests that $\phi_i$ should also be constant, in which case the second term vanishes by itself. Hence we simply have to balance the first term and the third one, which simplifies upon using the Killing spinor equation  \eqref{eqn:Killing}.

Recalling that $\eps^{ab} = \eps_i (\sigma_i)^{ab}$ is traceless Hermitian, we can take the $\grSU(2)_R$ trace of $\delta \lambda^{ab}$ which implies that $\eps_i \phi_i = 0$, i.e. the non-trivial supersymmetry variation parameter has to be orthogonal to the background scalar in the R-symmetry directions.

Given this restriction, we can now contract $\delta \lambda^{ab}$ with $(\sigma_i)_{ab}$ and find
\be
\tfrac{i}{2} \levi^{mnl} F_{mn} \gamma_l \, \eps_i + \levi_{i j k} \phi_j \gamma^\tau \eps_k = 0 \; .
\ee
Thus the magnitudes of the scalar background and gauge fields are related, and picking an $\grSU(2)_R$ orientation we can choose e.g. $\half \levi^{mn\tau} F_{mn}  = - \eta\, \phi_3$, where $\eta = \pm 1$ distinguishes BPS from anti-BPS monopoles. In this case $\eps_3=0$ and the remaining supersymmetry parameters have to satisfy\footnote{In Euclidean space we treat $\eps_1 \pm i \eps_2$ as two independent supersymmetry parameters.} $\eps_1 - i \eta \, \eps_2 = 0$.

Let us now turn to the terms of order $\tg$. Following the same lines of reasoning, they imply
\be \label{eqn:PhiXX}
 \phi^b_a =
   - \tfrac{1}{\kappa} \bigbrk{ X^b X^\dagger_a - \half \delta^b_a X X^\dagger } \quad \Leftrightarrow \quad \phi_i  = - \tfrac{1}{2 \kappa} X \sigma_i X^\dagger \; .
\ee
It is evident that if we choose $\hat{A} = A$ and $\hat{\phi}_i = - \phi_i$ the variations $\delta \hat{\lambda}^{ab}$  \eqref{eqn:SusyLambdaHat} will vanish also in the same manner, provided that
\be \label{eqn:PhiHatXX}
 \hat{\phi}^b_a =
   \tfrac{1}{\kappa} \bigbrk{ X^\dagger_a X^b - \half \delta^b_a X^\dagger X } \quad \Leftrightarrow \quad \hat{\phi}_i  = \tfrac{1}{2 \kappa} X^\dagger  \sigma_i X \; .
\ee
This leaves the supersymmetry transformations of the remaining fermions, $\delta \xi^{Aa}$ and its complex conjugate  \eqref{eqn:SusyXi}, to be verified. Given that $\hat{\phi}_i = - \phi_i$ causes the last two terms to cancel, they simply fix the functional dependence of the bifundamental scalars to be $X^{A1} \sim \exp{(-\eta\, \tau /2)}$ and $X^{A2} \sim \exp{(\eta\, \tau /2)}$.\footnote{And similarly $X^\dagger_{A1} \sim \exp{(\eta\, \tau /2)}$ and $X^\dagger_{A2} \sim \exp{(-\eta\, \tau /2)}$. While it may appear unusual that $X$ and $X^\dagger$ are not complex conjugates of each other, this is simply an artefact of the Euclidean signature of spacetime. After a suitable Wick rotation they would evidently be conjugate in the usual sense.} Then all of the $\delta \xi^{Aa}$ and $\delta \xi^\dagger_{Aa}$ vanish either by virtue of this particular $\tau$-dependence, which is consistent with the equation of motion for $X$
\be
  \deriD^2 X^{Aa}  - \quarter X^{Aa} = 0 \; ,
\ee
or because of a vanishing variation parameter.

Fixing the coefficients of the $X$ fields such that \eqref{eqn:PhiXX} and \eqref{eqn:PhiHatXX} are satisfied, the above BPS conditions are of course also consistent with the remaining equations of motion. In particular, those for the gauge fields, given that the Dirac monopole potential
satisfies $\deriD_n \bigbrk{ \tfrac{1}{\tg^2} F^{mn} } =  \deriD_n \bigbrk{ \tfrac{1}{\tg^2} \hat{F}^{mn} } = 0$, reduce to
\be \label{eqn:FXDX}
  \kappa \, \levi^{mnk} F_{nk}       \eq X \deriD^m X^\dagger - \deriD^m X X^\dagger   \; , \\
  \kappa \, \levi^{mnk} \hat{F}_{nk} \eq \deriD^m X^\dagger X - X^\dagger \deriD^m X  \nn \; .
 \ee

In summary, a convenient choice of classical (anti-)BPS solution is given by
\be \label{eqn:BPS-fixed}
  A = \hat{A} = \frac{H}{2} (\pm 1 - \cos\theta) \ext\varphi
  \comma
  \phi_i = - \hat{\phi}_i = - \eta \frac{H}{2} \delta_{i3}  \; ,
\ee
(where the upper sign holds on the northern and the lower one on the southern hemisphere), supplemented by an appropriate $X$ expectation value chosen to satisfy \eqref{eqn:PhiXX}, \eqref{eqn:PhiHatXX} and \eqref{eqn:FXDX}, e.g. for positive semi-definite $H$
\be
X^{11} \eq Z^1 = \sqrt{H \kappa}\, e^{- \tau/2} \ , \hspace{5mm} X^\dagger_{11} = Z^\dagger_1 = \sqrt{H \kappa}\, e^{\tau/2}  \hspace{4mm} \mathrm{ for }\  \eta = 1\  \mathrm{ or } \\
X^{12} \eq W^{1\dagger} = \sqrt{H \kappa}\, e^{-\tau/2} \ , \hspace{2.2mm} X^\dagger_{12} = W_1 = \sqrt{H \kappa}\, e^{\tau/2}  \hspace{3.2mm} \mathrm{ for }\  \eta = -1 \; , \nn
\ee
with all other $X$'s vanishing. Note that the choice of $X$ expectation value breaks the flavor symmetry and here we have arbitrarily used the first flavor. This background is invariant under supersymmetry transformations with the parameter $\eps_1 + i \eta \eps_2$, and can be generalized to any preferred  $\grSU(2)_R$ orientation.

Our classical solution is BPS along the full RG flow for any value of $\tg$, which is an important prerequisite for our arguments. However, the actual calculation we wish to carry out will be performed in the far UV, and here the role of the $X$ expectation value is quite different from the expectation values of the adjoint scalars and gauge fields. If we were to do perturbation theory in $g$ around this background, we would be led to rescale the (quantum) fields $A$ and $\phi$ (and their hatted analogues) by a factor of $g$ (before carrying out the Weyl transformation), and thus $g$ sets the scale of quantum fluctuations. Since the background values of  $A$ and $\phi$ are of order unity, they are parametrically larger than these fluctuations in the UV, and thus can be treated classically.

The fluctuations of the bifundamental field $X$ do not suffer such a rescaling however, and both quantum excitations as well as the expectation value in our classical solution are of the same order of magnitude. Therefore, we shall not treat the $X$ fields as a classical background in the UV theory. They are to be though of as quantum excitations which dress up the monopole background, and thus we shall drop them for the purpose of describing the bare monopole operator. We keep in mind however, that a bare monopole operator is not gauge invariant, and will eventually have to be contracted with a number of basic fields appearing in the Lagrangian in order to form a gauge invariant operator. In this context it is interesting to note that the number of $X$ excitations corresponding to our classical solution is of order $H k$, which is related to the number of fundamental fields we expect to contract the bare monopole operator with.

Finally, we will need to generalize the monopole background to arbitrary, possibly $\tau$-dependent $\grSU(2)_R$ orientation $n_i$ and thus, to conclude this discussion, we collect the basic properties of the semi-classical BPS monopoles background we will make use of below.

\paragraph{BPS monopole background.} For the background to be BPS, we have to essentially identify the two gauge groups $\grU(N)$ and $\hat{\grU}(N)$, and turn on expectation values for the gauge fields and adjoint scalars $\phi_i$ and $\hat{\phi}_i$ given by
\be \label{eqn:BPS-background}
  A = \hat{A} = \frac{H}{2} (\pm 1 - \cos\theta) \ext\varphi
  \comma
  \phi_i = - \hat{\phi}_i = - \frac{H}{2} n_i(\tau)
  \; .
\ee
The monopole background is diagonal in the $\grU(N)$ gauge indices $H = \diag(q_1, \ldots, q_N)$ where $q_r\in\Integers$ are the $\grU(1)^N$ gauge charges. They determine how the monopole transforms under gauge transformations. Furthermore, the background is labeled by a unit $\grSU(2)_R$ vector $n_i$ which gives the direction of the scalar fields. This vector is a collective coordinate of the background \eqref{eqn:BPS-background} and spans the moduli space $\Sphere^2$.

%%%%%%%%%%%%%%%%%%%%%%%%%%%%%%%%%%%%%%%%%%%%%%%%%%%%%%%%%%%%%%%%%%%%%%%%%%%
%%%%%%%%%%%%%%%%%%%%%%%%%%%%%%%%%%%%%%%%%%%%%%%%%%%%%%%%%%%%%%%%%%%%%%%%%%%
\section{$\grU(1)_R$ charges from normal ordering}
\label{sec:U1}

In this section we compute the quantum corrections to the $\grU(1)_R$ charge which is preserved by the static background \eqref{eqn:BPS-background} with $n_i = \delta_{i3}$. These quantum corrections are due to fermions fluctuations and are encoded in the normal ordering constant of the $\grU(1)_R$ charge operator. Before going into the specifics of ABJM theory, we will discuss the computation for a toy example which is then simple to generalize.

%%%%%%%%%%%%%%%%%%%%%%%%%%%%%%%%%%%%%%%%%%%%%%%%%%%%%%%%%%%%%%%%%%%%%%%%%%%
\subsection{Prototype}
\label{sec:U1-prototype}

Let us consider a single fermion $\psi(\tau,\Omega)$ in an abelian gauge theory subject to the equation of motion
\be \label{eqn:toy-eom}
  \mbox{} \hspace{30mm} \slashed{\deriD} \psi + \frac{\eta}{2} q \psi = 0 \; , \hspace{15mm} (\eta=\pm1)
\ee
and compute the oscillator expansion of the charge
\be \label{eqn:toy-U1R-charge}
  Q = -i \int\!d\Omega\: \psi^\dagger \gamma^\tau \psi \; .
\ee
The Dirac operator $\slashed{\deriD} = \slashed{\nabla} + i \slashed{A}$ in \eqref{eqn:toy-eom} includes a monopole background of the kind \eqref{eqn:BPS-background} with magnetic charge $H \to q$. The mass term in \eqref{eqn:toy-eom} whose magnitude is proportional to $q$ plays the role of the coupling to the background scalar.  The two signs, $\eta=\pm1$, correspond to a BPS and an anti-BPS background, respectively.

\begin{figure}
\begin{center}
\subfigure[Dirac operator]{\label{fig:ev-Dirac}\includegraphics{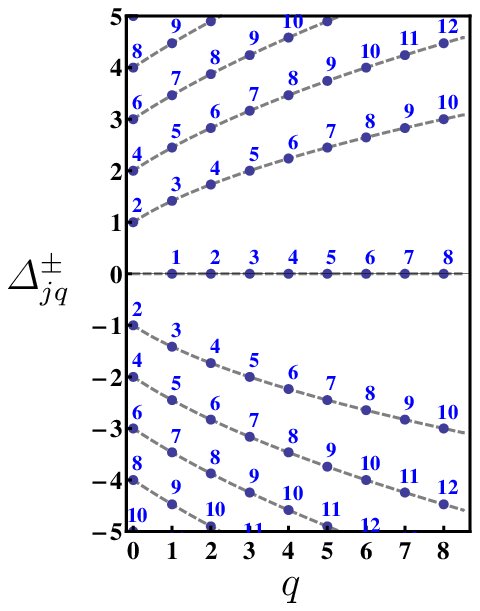}} \hspace{10mm}
\subfigure[BPS]{\label{fig:ev-BPS}\includegraphics{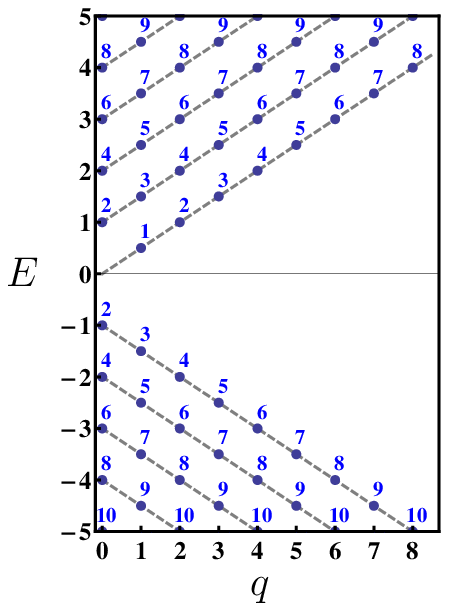}} \hspace{10mm}
\subfigure[anti-BPS]{\label{fig:ev-aBPS}\includegraphics{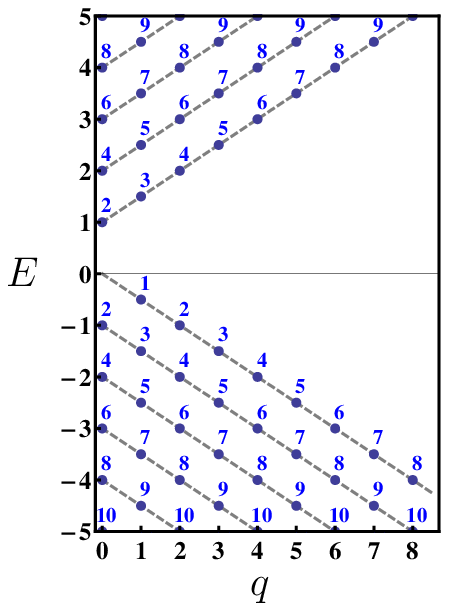}}
\end{center}
\caption{\textbf{\mathversion{bold} Eigenvalues of the Dirac operator on the $\Sphere^2$ and energy spectra for BPS and anti-BPS backgrounds.} The eigenvalues $\Delta^\pm_{jq} = \pm \half \sqrt{(2j+1)^2 - q^2}$ and $E = \pm E_j = \pm (j+\half)$ are parametrized by $j=\frac{\abs{q}-1}{2}+p$ for $p=0,1,2,\ldots$. The dashed lines correspond to a fixed value $p$. The numbers next to the points denote the multiplicities of the corresponding eigenvalues. Note in particular that there are $\abs{q}$ zero modes of the Dirac operator for monopole charge $q$, which are lifted to non-zero unpaired modes when the background scalar is turned on.}
\label{fig:ev}
\end{figure}

The easiest way to solve \eqref{eqn:toy-eom} is to expand $\psi$ in monopole spinor harmonics which are eigenfunctions of the monopole Dirac operator on the sphere, $\slashed{\deriD}_S$. This operator is contained in the full operator in \eqref{eqn:toy-eom} simply as $\slashed{\deriD} = \gamma^\tau \partial_\tau + \slashed{\deriD}_S$, since the monopole potential does not have any component along $\tau$. We note the explicit form of the monopole spinor harmonics in \appref{sec:monopole-spinor-harmonics}. Here we only need their eigenvalues with multiplicities. For given magnetic charge $q$, the quantum numbers of the total angular momentum are given by
\be
j \eq \frac{\abs{q}-1}{2}+p \qquad \text{with} \qquad p = 0,1,2,\ldots \; , \\
m \eq -j,-j+1,\ldots,j \; ,
\ee
where the state $j = \frac{\abs{q}-1}{2}$ is absent for $q=0$. We denote the eigenfunctions by
\be
     \slashed{\deriD}_\Sphere \Bsi^0_{qm}     \eq 0
     \hspace{29mm} \mbox{for $j=\frac{\abs{q}-1}{2}$} \\
     \slashed{\deriD}_\Sphere \Bsi^\pm_{qjm} \eq i \Delta^\pm_{jq} \Bsi^\pm_{qjm}
     \hspace{15mm} \mbox{for $j=\frac{\abs{q}+1}{2}, \frac{\abs{q}+3}{2}, \ldots$}
\ee
with eigenvalues
\be
  \Delta^\pm_{jq} = \pm \half \sqrt{(2j+1)^2 - q^2} \; .
\ee
This spectrum is plotted in \figref{fig:ev-Dirac}. The $\abs{q}$ zero modes which exist for non-zero magnetic charge will be responsible for a shift of the quantized version of the charge \eqref{eqn:toy-U1R-charge}.

Now we are ready to write the harmonic expansion of the wave-function as
\be \label{eqn:harmonic-expansion}
  \psi(\tau,\Omega) = \sum_m \psi_m(\tau) \Bsi^0_{qm}(\Omega) + \sum_{jm\eps} \psi^\eps_{jm}(\tau) \Bsi^\eps_{qjm}(\Omega)
\ee
where $\eps=\pm1$. Plugging this expansion into the equation of motion \eqref{eqn:toy-eom} and using the property \eqref{eqn:msh-gammatau},\eqref{eqn:msh-gammatau-zero} and the orthogonality \eqref{eqn:msh-orthogonal} one finds
\be
  \dot{\psi}_m = - \eta \frac{\abs{q}}{2} \psi_m
  \comma
  \matr{c}{\dot{\psi}^+_{jm} \\ \dot{\psi}^-_{jm}} \eq \matr{cc}{0 & -i\Delta^- - \eta \frac{q}{2} \\ -i\Delta^+ - \eta \frac{q}{2} & 0 } \matr{c}{\psi^+_{jm} \\ \psi^-_{jm}}
  \; .
\ee
The solution is
\be
  \psi = \sum_m \Bigsbrk{ c_m \, u^0 \, e^{-\frac{\abs{q}}{2}\tau} + d^\dagger_m \, v^0 \, e^{\frac{\abs{q}}{2}\tau} } \Bsi^0_m
       + \sum_{jm\eps} \Bigsbrk{ c_{jm} \, u^\eps_j \, e^{-E_j\tau} + d^\dagger_{jm} \, v^\eps_j \, e^{E_j\tau} } \Bsi^\eps_{jm}
       \; ,
\ee
where the energies are given by $E_j=j+\half$. The wave-functions for the BPS case ($\eta=+1$) are given by
\begin{align}
  u^0 = 1 \; , \;\;
  v^0 = 0 \; , \;\;
  u^+_j = v^+_j = 1 \; , \;\;
  u^-_j = - v^-_j = \tfrac{1}{\sqrt{2}} \Bigbrk{ \tfrac{q}{2j+1} + i \sqrt{1-\bigbrk{\tfrac{q}{2j+1}}^2} } \; ,
\end{align}
and the ones for the anti-BPS case ($\eta=-1$) are obtained from this by exchanging $u^0 \leftrightarrow v^0$ and $u^\pm_j \leftrightarrow v^\mp_j$. The normalization of the wave-functions are such that we have
\be
  \acomm{\psi_\alpha(\tau,\Omega)}{\Pi^\beta(t,\Omega')} \eq \delta_\alpha^\beta \delta^{(2)}(\Omega-\Omega') \; , \\
  \acomm{c_{jm}}{c^\dagger_{j'm'}} \eq \delta_{jj'} \delta_{mm'} \; , \\
  \acomm{d_{jm}}{d^\dagger_{j'm'}} \eq \delta_{jj'} \delta_{mm'} \; ,
\ee
where $\Pi = -i\psi^\dagger \gamma^\tau$ is the canonically conjugate momentum.

The energy spectrum is plotted in \figref{fig:ev-BPS} and \figref{fig:ev-aBPS}. One observes that the zero modes of the Dirac operator have turned into ``unpaired states'', i.e. states for which there are no states with the opposite energy in the spectrum. There are $2j+1=\abs{q}$ unpaired states with energy $\eta\frac{\abs{q}}{2}$, which is positive for the BPS and negative for the anti-BPS case.

Now we compute the $\grU(1)_R$ charge using point splitting regularization as in \cite{Borokhov:2002cg}:
\be \label{eqn:U1R-def-point-splitting}
  Q(\beta) = -\iHalf \int\!d\Omega\: \Bigsbrk{
             \psi^\dagger\bigbrk{\tau+\tfrac{\beta}{2}} \, \gamma^\tau \, \psi\bigbrk{\tau-\tfrac{\beta}{2}}
           - \psi\bigbrk{\tau+\tfrac{\beta}{2}} \, \gamma^\tau \, \psi^\dagger\bigbrk{\tau-\tfrac{\beta}{2}}
           } \; .
\ee
where $\beta>0$ will be taken to zero in the end. Inserting the oscillator expansion\footnote{The oscillator expansion of $\psi^\dagger$ is the complex conjugate of that of $\psi$ where in addition the sign of $\tau$ is reversed.} and ordering the terms, we find the normal ordered piece
\be
  Q_1(\beta=0) = \sum_{jm} \Bigsbrk{ c^\dagger_{jm} c_{jm} - d^\dagger_{jm} d_{jm} }
\ee
and a normal ordering constant
\be
  Q_0(\beta) = - \Half \sum_{jm\eps} \Bigsbrk{ u^{\eps\dagger}_j u^\eps_j - v^{\eps\dagger}_j v^\eps_j } e^{-\beta E_j} \; ,
\ee
where in both sums we understand the zero modes with $j=\frac{\abs{q}-1}{2}$ to be included. For that value of $j$ there is no sum over $\eps=\pm1$ and $E_j = \frac{\abs{q}}{2}$. The normal ordering constant can be written in a concise form by noticing that $\sum_\eps u^{\eps\dagger}_j u^\eps_j = 1$ gives a contribution for every positive energy state and $\sum_\eps v^{\eps\dagger}_j v^\eps_j = 1$ one for every negative energy state. Hence we can write
\be
  Q_0(\beta) = -\Half \sum_{\mathrm{states}} \sign(E) \, e^{-\beta \abs{E}} \; ,
\ee
where the sum extends over all states in the spectrum\footnote{This formulae looks superficially different from that in \cite{Borokhov:2003yu} because we sum over \emph{all} states including their degeneracy, not just over different energies levels.} and vanishes if the spectrum is symmetric with respect to $E=0$. This is not the case due to the unpaired states. Instead we find
\be
  Q_0 = - \eta \frac{\abs{q}}{2} \; ,
\ee
where the factor $\abs{q}$ is the number of unpaired states (the degeneracy of the mode $j=\frac{\abs{q}-1}{2}$) and $\eta$ the sign of their energy. This normal ordering constant gives the $\grU(1)_R$ charge of the BPS or anti-BPS monopole background, in agreement with \cite{Borokhov:2002cg}. As it was also shown in \cite{Borokhov:2002cg}, the bosonic fields do not contribute to the induced charge because their spectrum is symmetric.

%%%%%%%%%%%%%%%%%%%%%%%%%%%%%%%%%%%%%%%%%%%%%%%%%%%%%%%%%%%%%%%%%%%%%%%%%%%
\subsection{Application to ${\superN}=3$ gauge theory}
\label{sec:U1-ABJM}

The above discussion is readily applied to the $\superN=3$ gauge theories with $N_f$ hyper multiplets. In place of the charge \eqref{eqn:toy-U1R-charge} we now have
\be \label{eqn:U1R-charge}
  Q \eq - i \int\!d\Omega\: \tr \Bigsbrk{ - \half \, \zeta^\dagger_A \gamma^\tau \zeta^A
                                          - \half \, \omega^{\dagger A} \gamma^\tau \omega_A
                                          + \chi_\sigma^\dagger \gamma^\tau \chi_\sigma
                                          + \hat{\chi}_\sigma^\dagger \gamma^\tau \hat{\chi}_\sigma } \; ,
\ee
see \eqref{eqn:U1R-Noether-current-fermions}, and the equation of motion \eqref{eqn:toy-eom} is replaced by
\begin{align} \label{eqn:eom-fluct-euclid}
  & \slashed{\deriD} \zeta^A    + \tfrac{\eta}{2} \comm{H}{\zeta^A} = 0 \; , &
  & \slashed{\deriD} \chisigma  + \tfrac{\eta}{2} \comm{H}{\chisigma} = 0 \; ,  &
  & \slashed{\deriD} \hchisigma + \tfrac{\eta}{2} \comm{H}{\hchisigma} = 0 \; ,  \\
  & \slashed{\deriD} \omega_A   + \tfrac{\eta}{2} \comm{H}{\omega_A} = 0 \; , &
  & \slashed{\deriD} \chiphi    + \tfrac{\eta}{2} \comm{H}{\chiphi} = 0 \; ,  &
  & \slashed{\deriD} \hchiphi   + \tfrac{\eta}{2} \comm{H}{\hchiphi} = 0 \; , \nn
\end{align}
which hold in the static background \eqref{eqn:BPS-fixed} and in the far UV where $\tg\to 0$. The only qualitative difference is that we are now dealing with a non-abelian theory. Since the background fields have the same $\grU(N)$ dependence for both gauge group factors, namely $H = \diag(q_1,\ldots,q_N)$, all fields effectively transform in the adjoint of the diagonal $\grU(N)_d \subset \grU(N)\times \hat{\grU}(N)$. This is apparent from the fact that we could write commutators in \eqref{eqn:eom-fluct-euclid}, and also the gauge field inside the Dirac operator acts via a commutator. Now, the key observation is
\be \label{eqn:key-observation}
  \comm{H}{\psi}_{rs} = q_r \delta_{rt} \, \psi_{ts} - \psi_{rt} \, q_t \delta_{ts} = ( q_r - q_s ) \psi_{rs} \; ,
\ee
where $\psi_{rs}$ is one of the $N\times N$ matrix elements. This allows us to consider all matrix elements separately, if we use
\be \label{eqn:eff-monople-charge}
  q \to q_{rs} \equiv q_r - q_s
\ee
as the effective monopole charge. This immediately implies that the $\grU(1)_R$ charge of the vacuum, i.e. the monopole background, is
\be \label{eqn:U1R-ABJM-result}
  Q_R^{\mathrm{mon}} \eq \sum_{r,s} \Bigsbrk{- \half \cdot N_f \cdot \lrbrk{ -\eta \tfrac{\abs{q_{rs}}}{2} }
                              - \half \cdot N_f \cdot \lrbrk{ -\eta \tfrac{\abs{q_{rs}}}{2} }
                              + 1 \cdot \lrbrk{ -\eta \tfrac{\abs{q_{rs}}}{2} }
                              + 1 \cdot \lrbrk{ -\eta \tfrac{\abs{q_{rs}}}{2} } }  \; ,
                              \hspace{10mm}\mbox{}
\ee
or simply
\be \label{eqn:induced-U1}
     Q_R^{\mathrm{mon}} \eq \eta \lrbrk{\frac{N_f}{2} - 1} \sum_{r,s=1}^N \abs{q_r-q_s} \; .
\ee
This is the generalization to $N_f$ flavors of (B.60) of \cite{Kim:2009wb} evaluated for $H=\hat{H}$.\footnote{The work of \cite{Kim:2009wb} also studied the more general case where $H\not=\hat{H}$.} The sum produces an answer at least of order $N$. For example, for the simplest monopole whose only non-vanishing label is $q_1=1$, and therefore transforms in the bifundamental of $\grU(N)\times \grU(N)$ for $k=1$, we find that the induced R-charge is $(N_f-2)(N-1)$. \\

For the BPS background where $\eta=+1$, the R-charge of the monopole is positive if $N_f > 2$. For $N_f=1$ the monopole R-charges are negative and the theory may not flow to a conformal limit \cite{Gaiotto:2008ak,Gaiotto:2009tk}. The BPS monopoles in ABJM theory have vanishing R-charge since there are two flavors, $N_f=2$. However, this does not mean that there are gauge invariant operators with vanishing R-charge because such operators necessarily involve matter fields in addition to the monopoles. In fact, in a strongly coupled theory it is generally impossible to separate the monopole part from the matter part; nevertheless, this is possible in the weakly coupled UV limit where the monopole part is semi-classical. \\

We finally remark that from a similar computation, where we normal order the flavor charge operator instead of the R-charge operator, one can see that the monopole does not carry any induced flavor representation. Using the mode expansion in the flavor charge
\be \label{eqn:flavor-charge}
  Q_{\mathrm{fl}}^n \sim \int\!d\Omega\: \tr \xi^\dagger_A (T^n)^A{}_B \gamma^\tau \xi^B
    = \int\!d\Omega\: \tr \Bigsbrk{ \zeta^\dagger_A (T^n)^A{}_B \gamma^\tau \zeta^B - \omega^{\dagger B} (T^n)^A{}_B \gamma^\tau \omega_A }
\ee
yields a vanishing normal ordering constant simply because the generators $(T^i)^A{}_B$ of $\grSU(N_f)$ are traceless. Note that considering the static background \eqref{eqn:BPS-fixed} which preserves only $\grU(1)_R\subset\grSU(2)_R$ is general enough for this argument, because flavor and R-charge are completely independent.

%%%%%%%%%%%%%%%%%%%%%%%%%%%%%%%%%%%%%%%%%%%%%%%%%%%%%%%%%%%%%%%%%%%%%%%%%%%
%%%%%%%%%%%%%%%%%%%%%%%%%%%%%%%%%%%%%%%%%%%%%%%%%%%%%%%%%%%%%%%%%%%%%%%%%%%
\section{$\grSU(2)_R$ charges from collective coordinate quantization}
\label{sec:SU2}

In this section we compute the $\grSU(2)_R$ charges of the BPS monopole operators by quantizing the collective coordinate $\vec{n}(\tau)$ of the corresponding class of classical monopole backgrounds \eqref{eqn:BPS-background}. The dynamics of the collective coordinate is governed by the interaction with the other fields of the theory. We take the influence of these interactions into account by computing the effective action $\Gamma(\vec{n})$. As explained in \secref{sec:heart} it is sufficient to carry out this computation in the UV limit of the theory where the Yang-Mills coupling $\tg$ goes to zero. In this limit precisely the interactions with the fermions survive. Thus the effective action is obtained by integrating out the fermions:
\be
  e^{-\Gamma(\vec{n})} = \int [d\xi^\dagger][d\xi][d\lambda][d\hat{\lambda}] \: e^{ -\Action } \; ,
\ee
where the relevant part of the action is
\be \label{eqn:relevant-action}
  \Action \eq \int\!d\tau\,d\Omega\: \tr\Bigsbrk{
       -i \xi^\dagger_{Aa} \slashed{\deriD} \xi^{Aa} - \ihalf \, n_i \, \xi^\dagger_{Aa} (\sigma_i)^a{}_b \comm{H}{\xi^{Ab}}
       \nl\hspace{22mm}
       +\ihalf \lambda_{ab} \slashed{\deriD} \lambda^{ab} - \ihalf \, n_i \, \lambda_{ab} (\sigma_i)^b{}_c \comm{H}{\lambda^{ac}}
       \nl[1mm]\hspace{22mm}
       +\ihalf \hat{\lambda}_{ab} \slashed{\deriD} \hat{\lambda}^{ab} - \ihalf \, n_i \, \hat{\lambda}_{ab} (\sigma_i)^b{}_c \comm{H}{\hat{\lambda}^{ac}}
  \, } \; .
\ee
Since the action is quadratic the path integral is ``simply'' a determinant, albeit a determinant of a matrix operator with very many indices. We have displayed explicitly the $\grSU(2)_{\mathrm{fl}}$ index ($A$) and the $\grSU(2)_R$ indices ($a,b,c$). Besides those, there are implicit $\grU(N)$ gauge indices which we denote by $\xi_{rs}$, $\lambda_{rs}$, etc.\ below. Then there is the spatial dependence of the fermions which we will trade for a set of mode indices by expanding the fields into harmonics on the sphere.

The good news is that the operator is fairly diagonal and couples only very few components together. For instance it is completely diagonal in the flavor and gauge indices, and couples at most two modes of the harmonic expansion. Thus we can perform the computation for a generic component and take the sum over the indices into account later.

%%%%%%%%%%%%%%%%%%%%%%%%%%%%%%%%%%%%%%%%%%%%%%%%%%%%%%%%%%%%%%%%%%%%%%%%%%%
\subsection{Prototype}
\label{sec:SU2-prototype}

We start the discussion with a quantum mechanical model where there is only a (Euclidean) time coordinate $\tau$. This example already exposes the essential point of the whole argument. The dependence of the fields on the angular coordinates will be taken into account in the next paragraph.

%%%%%%%%%%%%%%%%%%%%%%%%%%%%%%%%%%%%%%%%%%%%%%%%%%%%%%%%%%%%%%%%%%%%%%%%%%%
\paragraph{Quantum mechanics.} Let us consider one fermion $\psi^a(\tau)$ in the fundamental representation of $\grSU(2)_R$ as indicated by the index $a$, with the action\footnote{A similar model with a fermion in the $\rep{3}$ of $\grSU(2)$ was studied in \cite{Dusedau:1988ng}.}
\be \label{eqn:relevant-action-simple}
  \Action = \int\!d\tau\: \Bigsbrk{ -i \psi^\dagger_a \partial_\tau \psi^a - \ihalf \, q n_i(\tau) \, \psi^\dagger_a (\sigma_i)^a{}_b \psi^b } \; .
\ee
Since we have not included any spatial dependence in this example, there is no monopole gauge field either. The coupling of $\psi$ to the collective coordinate has been written as $q$ in anticipation of it becoming the monopole charge later.

Formally we find for the effective action\footnote{The relative sign between the two terms has superficially changed since $(\sigma_i)^a{}_b$ in \eqref{eqn:relevant-action-simple} are transposed Pauli matrices, which in \eqref{eqn:effective-action-from-fermion-determinant} we use non-transposed ones, $(\sigma_i)_a{}^b$.}
\be \label{eqn:effective-action-from-fermion-determinant}
  \Gamma(\vec{n}) = - \ln \det \Bigbrk{ i \partial_\tau - i \tfrac{q}{2} n_i(\tau) \sigma_i } \; .
\ee
Due to the unspecified $\tau$-dependence of $n_i$, we cannot evaluate this determinant exactly. However, since the collective coordinate is considered as being quasi-static, it is legitimate to do a derivative expansion. The general form of the effective action then is
\be
  \Gamma(\vec{n}) = \int\!d\tau \Bigsbrk{
  - V_{\mathrm{eff}}(\vec{n})
  + i \, \dot{n}_i A_i(\vec{n})
  + \half \dot{n}_i \dot{n}_j B_{ij}(\vec{n})
  + \ldots } \; .
\ee
Our aim is to find the function $A_i(\vec{n})$ by expanding \eqref{eqn:effective-action-from-fermion-determinant}. However, it is not immediately possible to expand \eqref{eqn:effective-action-from-fermion-determinant} in $\dot{n}_i$, as it does not contain $\dot{n}_i$ explicitly. The trick due to \cite{Fraser:1984zb} is to write
\be \label{eqn:fraser-trick}
  n_i(\tau) = \mathring{n}_i + \tilde{n}_i(\tau) \; ,
\ee
where $\mathring{n}_i$ is constant with $\mathring{n}^2 = 1$ and $\tilde{n}_i(\tau)$ a small ``fluctuation''. Then the form of the effective action to second order in fluctuations reads
\be \label{eqn:eff-action-expanded}
  \Gamma(\vec{n}) \eq \int\!d\tau \Bigsbrk{
  - V_{\mathrm{eff}}(\mathring{\vec{n}})
  - \tilde{n}_i \partial_i V_{\mathrm{eff}}(\mathring{\vec{n}})
  - \half \tilde{n}_i \tilde{n}_j \partial_i \partial_j V_{\mathrm{eff}}(\mathring{\vec{n}})
  \nl \hspace{20mm}
  + i \, \dot{\tilde{n}}_i A_i(\mathring{\vec{n}})
  + i \, \dot{\tilde{n}}_i \tilde{n}_j \partial_j A_i(\mathring{\vec{n}})
  + \half \dot{\tilde{n}}_i \dot{\tilde{n}}_j B_{ij}(\mathring{\vec{n}})
  + \ldots } \; .
\ee
The point of \eqref{eqn:fraser-trick} is that we can now expand \eqref{eqn:effective-action-from-fermion-determinant} in powers of $\tilde{n}_i$, some of which will come with $\tau$-derivatives, and compare the result with the general expression \eqref{eqn:eff-action-expanded}. We cannot determine $A_i$ from the term $\dot{\tilde{n}}_i A_i(\mathring{\vec{n}})$ as this is a total derivative and hence will not show up in the expansion of \eqref{eqn:effective-action-from-fermion-determinant}. Therefore we will focus on the next term, the unique term with two powers of $\tilde{n}$ and one $\tau$-derivative
\be \label{eqn:eff-action-2-1}
  \Gamma_{(2,1)}(\vec{n}) \eq i \int\!d\tau\: \dot{\tilde{n}}_i \tilde{n}_j \partial_j A_i(\mathring{\vec{n}})
                             = -\iHalf \int\!d\tau\: \dot{\tilde{n}}_i \tilde{n}_j
                                 \, \bigbrk{ \partial_i A_j(\mathring{\vec{n}}) - \partial_j A_i(\mathring{\vec{n}}) } \; .
\ee

Now we start from \eqref{eqn:effective-action-from-fermion-determinant} using \eqref{eqn:fraser-trick}
\be
  \Gamma(\vec{n}) \eq - \tr \ln \Bigbrk{ i \partial_\tau - i \mathring{\backslashed{m}} - i \tilde{\backslashed{m}} }
                  =   - \tr \ln \Bigbrk{ i \partial_\tau - i \mathring{\backslashed{m}} }
                      - \tr \ln \Bigbrk{ \unit - \tfrac{1}{\partial_\tau - \mathring{\backslashed{m}}} \tilde{\backslashed{m}} }
  \; ,
\ee
where we have introduced the shorthands $\mathring{m}_i = \frac{q}{2} \mathring{n}_i$, $\tilde{m}_i = \frac{q}{2} \tilde{n}_i$, and $\backslashed{m} \equiv m_i \sigma_i$. We isolate the term with two powers of $\tilde{n}_i \sim \tilde{m}_i$ by expanding the logarithm:
\be \label{eqn:Gamma2-simple}
  \Gamma_{(2)}(\vec{n}) \eq \Half \tr \biggsbrk{
     \frac{\partial_\tau + \mathring{\backslashed{m}}}{\partial_\tau^2 - \mathring{m}^2} \, \tilde{\backslashed{m}} \,
     \frac{\partial_\tau + \mathring{\backslashed{m}}}{\partial_\tau^2 - \mathring{m}^2} \, \tilde{\backslashed{m}} } \; .
\ee
Next we move all derivatives to the right using
\be
  \frac{1}{\partial^2 - \mathring{m}^2} \phi = \sum_{k=0}^\infty (-1)^k \underbrace{[\partial^2,[\partial^2,\ldots,[\partial^2}_{k},\phi]\ldots]] \frac{1}{(\partial^2 - \mathring{m}^2)^{k+1}}
  \; ,
\ee
and perform the trace over $\grSU(2)_R$ indices using
\be
  \tr \sigma_i \sigma_j = 2 \delta_{ij}
  \comma
  \tr \sigma_i \sigma_j \sigma_k = 2 i \levi_{ijk}
  \comma
  \tr \sigma_i \sigma_j \sigma_k \sigma_l = 2 \delta_{ij}\delta_{kl} - 2 \delta_{ik}\delta_{jl} + 2 \delta_{il}\delta_{jk}
  \; .
\ee
We find for the terms which contain exactly one derivative
\be \label{eqn:Gamma21-simple}
  \Gamma_{(2,1)}(\vec{n}) \eq
                       - 3 \tr \dot{\tilde{m}}_i \tilde{m}_i \frac{\partial_\tau (\partial_\tau^2 + \mathring{m}^2)}{(\partial_\tau^2 - \mathring{m}^2)^3}
                       - i \tr \levi_{ijk} \dot{\tilde{m}}_i \tilde{m}_j \mathring{m}_k \frac{1}{(\partial_\tau^2 - \mathring{m}^2)^2} \nl
                       - 6 \tr \bigbrk{ 2 \dot{\tilde{m}}_i \tilde{m}_j \mathring{m}_i \mathring{m}_j
                                        - \dot{\tilde{m}}_j \tilde{m}_j \mathring{m}_i \mathring{m}_i }
                                        \frac{\partial_\tau}{(\partial_\tau^2 - \mathring{m}^2)^3}
                       \; .
\ee
Now that the coordinate dependent part is separated from the derivatives it is easy to evaluate the functional trace: it leads to one integral over $\tau$ and one over the energy $\omega$, where $\partial_\tau \to - i \omega$. All but the second term will lead to an integral over a total $\tau$-derivative and therefore can be dropped. We are left with only the second term which becomes
\be \label{eqn:gamma21-result-simple}
  \Gamma_{(2,1)}(\vec{n})
    \eq -i \int\!d\tau\: \levi_{ijk} \dot{\tilde{m}}_i \tilde{m}_j \mathring{m}_k \int\!\frac{d\omega}{2\pi}\: \frac{1}{(\omega^2+\mathring{m}^2)^2}
    = -\frac{i}{4} \sign(q) \int\!d\tau\: \levi_{ijk} \dot{\tilde{n}}_i \tilde{n}_j \frac{\mathring{n}_k}{\abs{\mathring{\vec{n}}}{}^3} \; .
    \hspace{8mm} \mbox{}
\ee
Comparing this to \eqref{eqn:eff-action-2-1}, we read off
\be \label{eqn:monopole-potential-on-moduli-space}
  \partial_i A_j(\vec{n}) - \partial_j A_i(\vec{n}) = \frac{\sign(q)}{2} \, \levi_{ijk} \frac{n_k}{\abs{\vec{n}}{}^3} \; .
\ee
This expression is recognized as the field strength for a magnetic monopole with charge $\frac{\sign(q)}{2}$ and hence $A_i(\vec{n})$ is the corresponding gauge potential. We should stress that this monopole has nothing to do with the original monopole background of ABJM theory which we set out to study. In fact, the action \eqref{eqn:relevant-action-simple} does not contain any monopole background for $\psi$. The monopole potential we are finding here lives on the space spanned by $\vec{n}$, which just happens to be the moduli space of supersymmetric monopoles in $\superN=3$ gauge theory.

At any rate what we have found is that the effect of the fermion $\psi$ is to induce a Wess-Zumino term in the effective action for the collective coordinate. In other words, the dynamics of the collective coordinate $\vec{n}$ is the same as that of a point particle on a sphere with a magnetic monopole at its center. The coefficient of the Wess-Zumino term is the product of the electric charge of the point particle and the magnetic charge of the monopole. This analogy immediately implies that the allowed $\grSU(2)_R$ representations for the quantized collective coordinate, and hence the ABJM monopoles, are bounded from below by the Wess-Zumino coefficient. Thus, if we want to have monopole operators in the singlet representation, we will have to find that the Wess-Zumino term cancels from the effective action when all contributions are included. This is what will indeed happen for ABJM theory, but not in general.

%%%%%%%%%%%%%%%%%%%%%%%%%%%%%%%%%%%%%%%%%%%%%%%%%%%%%%%%%%%%%%%%%%%%%%%%%%%
\paragraph{Spatial dependence.} We reinstate the dependence of $\psi$ on the coordinates of the sphere and consider the case
\be \label{eqn:relevant-action-simple-spatial}
  \Action \eq \int\!d\tau\,d\Omega\: \Bigsbrk{
       -i \psi^\dagger_a \slashed{\deriD} \psi^a - \ihalf \, q n_i(\tau) \, \psi^\dagger_a (\sigma_i)^a{}_b \psi^b
  } \; ,
\ee
which contains a monopole background with magnetic charge $q$ but is still abelian. The easiest way to deal with the spatial dependence is to expand $\psi(\tau,\Omega)$ into monopole spinor harmonics and perform the $\Sphere^2$-integration in the action. All relevant properties of these harmonics have already been discussed in \secref{sec:U1-prototype}, see also \appref{sec:monopole-spinor-harmonics}.

The expansion of $\psi$ is the same as in \secref{sec:U1}, eq.\ \eqref{eqn:harmonic-expansion}. Since $\vec{n}(\tau)$ is constant on the sphere, the orthogonality of the monopole harmonics \eqref{eqn:msh-orthogonal} implies that different $(jm)$-modes of $\psi$ do not couple to each other. The only coupling is between the $\pm$-modes which is due to the property \eqref{eqn:msh-gammatau}. Indeed the action for the modes becomes
\be \label{eqn:action-mode-expansion}
  \Action \eq \sum_{m} \int\!d\tau\: \Bigsbrk{ -i \, \sign(q) \, \psi^\dagger_m \partial_\tau \psi_m
                                    - \ihalf \, q n_i \, \psi^\dagger_m \sigma_i \psi_m } \nl
          + \sum_{jm\eps} \int\!d\tau\: \Bigsbrk{ -i \psi^{\eps\dagger}_{jm} \partial_\tau \psi^\eps_{jm}
                                    + \Delta^\eps_{jq} \, \psi^{-\eps\dagger}_{jm} \psi^\eps_{jm}
                                    -\ihalf \, q n_i \, \psi^{-\eps\dagger}_{jm} \sigma_i \psi^\eps_{jm} } \; .
\ee
This decoupling means that we can compute the contribution to the effective action for each pair $(jm)$ individually. One such term in the zero-mode sector, $j=\frac{\abs{q}-1}{2}$, is essentially the previously considered case \eqref{eqn:relevant-action-simple}. The additional $\sign(q)$ in \eqref{eqn:action-mode-expansion}, which originates from \eqref{eqn:msh-gammatau-zero}, removes the $\sign(q)$ from the result \eqref{eqn:gamma21-result-simple}. The sum over $m$ introduces a factor of $2j+1=\abs{q}$. This is how the Wess-Zumino term acquires the dependence on the monopole charge. Thus the Wess-Zumino potential is
\be \label{eqn:WZ-potential}
  \partial_i A_j(\vec{n}) - \partial_j A_i(\vec{n}) = \frac{\abs{q}}{2} \, \levi_{ijk} \frac{n_k}{\abs{\vec{n}}{}^3} \; .
\ee

Now we turn to the non-zero-mode sector. It will turn out that there is no contribution to the effective action from this sector, and that \eqref{eqn:WZ-potential} is the final result. In the following we think of $j$ and $m$ as fixed to some values and suppress these labels. Due to the coupling between the $\pm$-modes, we now have a $2\times2$ matrix in ``mode space'' on top of the matrix structure that mixes the two components of the $\grSU(2)_R$ doublet:
\be
  \Gamma(\vec{n}) = - \ln \det \matr{cc}{
     i \partial_\tau & -\Delta^- - i \backslashed{m} \\
     -\Delta^+ - i \backslashed{m} & i \partial_\tau
   } \; .
\ee
The generalization of \eqref{eqn:Gamma2-simple} is
\be
  \Gamma_{(2)}(\vec{n}) \eq \Half \tr \lrsbrk{
     \raisebox{-4mm}{$\displaystyle \frac{\matr{cc}{ -\partial_\tau & -i\Delta-\mathring{\backslashed{m}} \\ i\Delta-\mathring{\backslashed{m}} & -\partial_\tau } }{\partial^2_\tau - \Delta^2 - \mathring{m}^2 }$} \, \matr{cc}{0 & \tilde{\backslashed{m}} \\ \tilde{\backslashed{m}} & 0 } }^2
\ee
where $\Delta \equiv \Delta^+ = -\Delta^-$. Evaluating this expression analogously to the previous case yields that the term $\Gamma_{(2,1)}(\vec{n})$ is zero (up to surface terms), i.e.\ the non-zero modes do not contribute to the Wess-Zumino term.

Recall that in the computation of the $\grU(1)_R$ charge of the monopole operator in \secref{sec:U1}, we also found that the non-zero modes did not contribute. There the cancellation occurred between states of equal but opposite energy. In order to demonstrate explicitly that the same mechanism is at work here, too, we pretend that $\Delta^+$ and $\Delta^-$ are unrelated for the time being and repeat the computation. Without presenting any of the lengthy intermediate results, we arrive at
\be \label{eqn:gamma21-result-general}
  \Gamma_{(2,1)}(\vec{n}) \eq
       \int\!d\tau\int\!\frac{d\omega}{2\pi}\: \frac{
            -4i \levi_{ijk} \dot{\tilde{m}}_i \tilde{m}_j \mathring{m}_k \,
            (\Delta^+ + \Delta^-) (\omega^2 + \Delta^+\Delta^-) \, \omega
          }{
            \bigsbrk{\omega^4+2(\Delta^+\Delta^- - \mathring{m}^2) \omega^2
             + (\Delta^+\Delta^-)^2 + \bigbrk{(\Delta^+)^2 + (\Delta^-)^2} \mathring{m}^2 + \mathring{m}^4 }^2 } \; . \nl
\ee
Indeed we see that the vanishing is due to the pairing of eigenvalues, $\Delta^+ = - \Delta^-$.

%%%%%%%%%%%%%%%%%%%%%%%%%%%%%%%%%%%%%%%%%%%%%%%%%%%%%%%%%%%%%%%%%%%%%%%%%%%
\subsection{Application to $\superN=3$ gauge theory}
\label{sec:SU2-ABJM}

Having obtained the result \eqref{eqn:WZ-potential} for the prototype action \eqref{eqn:relevant-action-simple-spatial} it is a simple matter to specialize to ABJM theory and its $\superN=3$ UV completion. All that needs to be done is to include the gauge indices and sum over the field content.

%%%%%%%%%%%%%%%%%%%%%%%%%%%%%%%%%%%%%%%%%%%%%%%%%%%%%%%%%%%%%%%%%%%%%%%%%%%
\paragraph{Gauge structure.} The non-abelian nature of the theory is taken care of just as in \secref{sec:U1-ABJM}. From \eqref{eqn:key-observation} it is clear that the action is diagonal in gauge indices and therefore every matrix element contributes independently from the others to the effective action. The effective monopole charge that the matrix element $\psi_{rs}$ experiences is given by
\be
  q_{rs} \equiv q_r - q_s \; .
\ee
Hence the non-abelian result is obtained from the abelian one by the replacement
\be
  \Gamma_q(\vec{n}) \to \sum_{r,s=1}^N \Gamma_{q_{rs}}(\vec{n}) \; .
\ee
We will see that the sum over $r,s$ will factorize, because all fermions transform effectively in the same representation. Therefore it will be convenient to define the total charge
\be
q_{\mathrm{tot}} = \sum_{r,s=1}^N \abs{q_{rs}} = 2 \sum_{r>s} \abs{q_r - q_s} \; .
\ee

%%%%%%%%%%%%%%%%%%%%%%%%%%%%%%%%%%%%%%%%%%%%%%%%%%%%%%%%%%%%%%%%%%%%%%%%%%%
\paragraph{Hyper multiplet fermions.} Upon using \eqref{eqn:key-observation}, the action for $\xi^A$ is just $N_f$ copies of the prototype \eqref{eqn:relevant-action-simple-spatial}. Therefore we can immediately write down the total contribution from the hyper multiplet fermions
\be
  \partial_i A_j(\vec{n}) - \partial_j A_i(\vec{n}) = N_f \, \frac{q_{\mathrm{tot}}}{2} \, \levi_{ijk} \frac{n_k}{\abs{\vec{n}}{}^3} \; .
\ee

%%%%%%%%%%%%%%%%%%%%%%%%%%%%%%%%%%%%%%%%%%%%%%%%%%%%%%%%%%%%%%%%%%%%%%%%%%%
\paragraph{Vector multiplet fermions.} The computation for $\lambda$ and $\hat{\lambda}$ reduces to \eqref{eqn:relevant-action-simple-spatial} as well, but we have to be careful not to over-count the degrees of freedom. Due to the relations \eqref{eqn:relations} there are only two independent (complex) components. We choose $\lambda^{11} \sim \chisigma$ and $\lambda^{12} \sim \chi_\phi^\dagger$ as the independent components and denote their complex conjugates by
\be
  \lambda^\dagger_{11} \equiv (\lambda^{11})^*
  \comma
  \lambda^\dagger_{12} \equiv (\lambda^{12})^*
  \; ,
\ee
and similarly for $\hat{\lambda}$. When expressed in terms of these fields, the action reads
\be
  \Action \eq \int\!d\tau\,d\Omega\: \tr\Bigsbrk{
       -i \lambda^\dagger_{1a} \slashed{\deriD} \lambda^{1a} + \ihalf \, n_i \, \lambda^\dagger_{1a} (\sigma_i)^a{}_b \comm{H}{\lambda^{1b}}
       \nl[1mm]\hspace{22mm}
       -i \hat{\lambda}^\dagger_{1a} \slashed{\deriD} \hat{\lambda}^{1a} + \ihalf \, n_i \, \hat{\lambda}^\dagger_{1a} (\sigma_i)^a{}_b \comm{H}{\hat{\lambda}^{1b}}
  \, } \; .
\ee
This is nothing but the action for $\xi$ with reversed sign of the interaction, cf.\ \eqref{eqn:relevant-action}. Thus the contribution from the vector multiplet fermions is
\be
  \partial_i A_j(\vec{n}) - \partial_j A_i(\vec{n}) = -2\times \frac{q_{\mathrm{tot}}}{2} \, \levi_{ijk} \frac{n_k}{\abs{\vec{n}}{}^3} \; ,
\ee
where the factor of $2$ arises because we have $\lambda$ and $\hat{\lambda}$.

%%%%%%%%%%%%%%%%%%%%%%%%%%%%%%%%%%%%%%%%%%%%%%%%%%%%%%%%%%%%%%%%%%%%%%%%%%%
\paragraph{Collective coordinate quantization.}

Adding all contributions together, the total effective action for the collective coordinate is given by
\be \label{eqn:collcoordaction}
  \Gamma(\vec{n}) = \int\!d\tau\: \Bigsbrk{
       \half M \dot{\vec{n}}^2
       + i \, \vec{A}(\vec{n})\cdot\dot{\vec{n}}
       + \lambda (\vec{n}^2 - 1)
     } \; ,
\ee
where the kinetic term is simply the sum of the kinetic terms for $\phi$ and $\hat{\phi}$. The last term is a Lagrange multiplier term which enforces the constraint that the modulus of $\vec{n}$ is fixed to one. This actions describes a particle with unit electric charge and large mass (as $\tg\to0$ in the UV)
\be
  M \eq M(\tau) = \frac{1}{\tg^2} \tr H^2 = g^{-2} e^{-2\tau} \sum_{r} q_r^2
\ee
on a sphere surrounding a magnetic monopole with charge
\be \label{eqn:induced-mag-charge}
  h \eq (N_f - 2) \, q_{\mathrm{tot}} = (N_f - 2) \sum_{r,s} \abs{q_r - q_s} \; ,
\ee
and field strength $\vec{B} = \vec{\nabla}\times\vec{A} = \frac{h}{2} \vec{n}$.

This is the Euclidean version of the system discussed in \secref{sec:heart} with the modification that now $M$ depends on time. Still, the conserved angular momentum is given by
\be \label{eqn:conserved-angular-momentum}
  \vec{L} = i M \, \vec{n}\times \dot{\vec{n}} - \frac{h}{2} \, \vec{n} \; ,
\ee
and its quantized values are the $\grSU(2)_R$ charges of the monopole operator described by the BPS background \eqref{eqn:BPS-background}. Due to the second term in \eqref{eqn:conserved-angular-momentum}, the smallest possible $\grSU(2)_R$ representation has spin
\be \label{eqn:induced-SU2}
  l = \frac{\abs{h}}{2} = \lrabs{\frac{N_f}{2} - 1} \sum_{r,s} \abs{q_r - q_s}
\ee
and dimension $2l+1=\abs{h}+1$.

A monopole in the singlet representation and hence with vanishing IR dimension, is only possible if $h=0$. One way of achieving this is to set all fluxes $q_r$ equal. This corresponds to a monopole in the diagonal $\grU(1)_d\subset\grU(N)\times\grU(N)$ of the gauge group which decouples from the matter fields. Another way to have singlet monopoles is to consider $N_f=2$ hyper multiplets, which is the field content of ABJM theory. This is true regardless of how the fluxes $q_r$ are distributed inside the total $\grU(N)_d$ flux $H$. \\

It is instructive to study the behavior of the collective coordinate wave-function as a function of $\tau$. We solve the Euclidean Schr\"odinger equation, $-\partial_\tau \psi = H \psi$, with the Hamiltonian
\be
 H = \frac{(\vec{p}-i\vec{A})^2}{2M} = \frac{\vec{L}^2-h^2/4}{2M}
\ee
derived from \eqref{eqn:collcoordaction}. This UV Hamiltonian is only valid for $\tau\to-\infty$; its eigenfunctions go to zero as $\exp(-\mathrm{const}\, e^{2\tau})$ at large $\tau$ which does not allow us to read off a well-defined energy. This behavior is corrected by including the two-derivative terms in the effective action due to the interactions of the collective coordinate with the unpaired fermion modes. These terms can be combined with the kinetic term in \eqref{eqn:collcoordaction} and the net effect is that the mass gets shifted to
\be
  \mu(\tau) = M(\tau) + \half \; .
\ee
With the ansatz $\psi(\tau,\vec{n}) = f(\tau)Y_{hlm}(\vec{n})$, we find
\be
 f \sim \bigbrk{\sqrt{\mu} \, e^\tau}^{-\omega}
 \sim \begin{cases}
   \mathrm{const.} & \text{for $\tau \to -\infty$ (UV)} \\
   e^{-\omega\tau} & \text{for $\tau \to +\infty$ (IR)}
 \end{cases}
\ee
where $\omega = l(l+1) - \frac{h^2}{4}$. In particular, for the lowest allowed value of the angular momentum, $l=\frac{\abs{h}}{2}$, the wave-function falls off as $\exp\bigbrk{-\frac{\abs{h}}{2}\tau}$. According to the operator state correspondence, a local operator inserted at the origin of $\Reals^3$ corresponds to a wave-function which is peaked in the UV ($\tau\to-\infty$), and decays at large $\tau$ as $e^{-\Delta \tau}$. We note that the wave-function we have found obeys these properties, and read off the conformal dimension of the corresponding monopole operator, $\Delta=\frac{\abs{h}}{2}$. It is interesting to see that this dimension is equal to the R-charge of the operator, just as it should be for BPS states. There could have been further corrections coming from higher loops, and the fact that one should include the bifundamental excitations in the infrared. Nevertheless ignoring all these complications, we still find the desired $\tau\to\infty$ asymptotics of the wave-function.

%%%%%%%%%%%%%%%%%%%%%%%%%%%%%%%%%%%%%%%%%%%%%%%%%%%%%%%%%%%%%%%%%%%%%%%%%%%
%%%%%%%%%%%%%%%%%%%%%%%%%%%%%%%%%%%%%%%%%%%%%%%%%%%%%%%%%%%%%%%%%%%%%%%%%%%
\section{Conclusions and outlook}
\label{sec:conclusions}

In this paper we calculated the global charges and dimensions of monopole operators in certain three-dimensional ${\superN}=3$ supersymmetric Yang-Mills Chern-Simons theories. This is the smallest amount of supersymmetry leading to a non-abelian R-symmetry which was crucial for our argument, because the $\grSU(2)_R$ spin of a monopole operator cannot change along an RG flow. This allowed us to find the exact charges from a one-loop calculation in the weakly coupled UV limit of the gauge theory. \\

In the far UV the monopole operator was adequately described by a classical Dirac monopole background for the gauge fields. For the description of BPS monopoles the background needs to be supersymmetric which required us to also turn on a classical background for the adjoint scalar fields. \\

In \secref{sec:U1} we considered a static scalar background. Since such a background breaks the R-symmetry from $\grSU(2)_R$ to $\grU(1)_R$, we could only determine the abelian charge of the monopole in this case. Using the methods developed in \cite{Borokhov:2002cg}, we found that fermionic fluctuations around the background induce a $\grU(1)_R$ charge of the monopole proportional to the R-charges of the fermions times their magnetic coupling to the background. Our complete formula \eqref{eqn:induced-U1} for the R-charge in $\grU(N)\times \grU(N)$ gauge theory coupled to $N_f$ hyper multiplets in the bifundamental representation is consistent with the proposal made in \cite{Gaiotto:2008ak,Gaiotto:2009tk}. \\

However, knowing the $\grU(1)_R$ charges at small Yang-Mills coupling is in general not enough as this quantity is not protected and one cannot make any reliable statement about their IR values. In \cite{Borokhov:2002cg} the computation was performed directly in the IR limit (of SQED) which was possible by assuming a large number of flavors. Here we could not resort to this trick because we wanted to keep the number of flavors arbitrary. However, we can make use of the $\superN=3$ supersymmetry and compute non-abelian R-charges which \emph{are} protected. They follow from quantization of the $\grSU(2)/\grU(1)$ collective coordinate of the background. By calculating the fermionic determinants which induce a Wess-Zumino term in the effective action of the collective coordinate, we demonstrated in \secref{sec:SU2} that the smallest allowed $\grSU(2)_R$ representation is given by \eqref{eqn:induced-SU2}. The largest $\grU(1)_R$ charge within this representation coincides with our findings in \secref{sec:U1}. \\

%\remark{Do we want this here?}
Note that the induced R-charge of the monopole is entirely due to the fermions of the theory. In the normal ordering computation, \secref{sec:U1}, the reason why the bosons do not contribute is because their spectrum is symmetric with respect to zero and therefore the states with positive energy cancel the effect of those with negative energy. In the collective coordinate computation, \secref{sec:SU2}, this follows from the fact that the coupling between the bosons and the collective coordinate goes to zero in the UV. \\

After the theory has flown to the superconformal Chern-Simons fixed point, we can use these results to argue that the contribution of the monopole operator to gauge-invariant operator dimensions is given by \eqref{eqn:induced-U1}, as long as it is non-negative (when it is negative, a conventional fixed point does not exist). For ABJM theory, where $N_f=2$, this contribution vanishes which is crucial for matching the spectrum with supergravity on $AdS_4\times S^7/\Integers_k$ and for the supersymmetry enhancement to $\superN=8$. \\

Since the gauginos make a crucial negative contribution to the R-charge, and they are not even dynamical in the IR Chern-Simons theory, it is not clear how to carry out this calculation reliably without appealing to the UV theory containing the Yang-Mills term. Luckily, there are various other theories to which the method of starting with a weakly coupled UV theory can be applied. One obvious example is to consider quiver theories with more than two $\grU(N)$ gauge groups and bifundamental hyper multiplets. These Yang-Mills Chern-Simons theories can flow in the IR to $\superN=3$ superconformal fixed points; some of them have M-theory $AdS_4$ duals found by Jafferis and Tomasiello \cite{Jafferis:2008qz}. \\

Determination of monopole operator dimensions poses more of a challenge in $\superN=2$ superconformal Chern-Simons theories,\footnote{I.R.K. is grateful to D.\ Jafferis and S.\ Pufu for very useful discussions on this issue.} since we cannot rely on a non-abelian R-symmetry. Nevertheless, it should again be possible to define some of these theories via flow from weakly coupled Yang-Mills Chern-Simons theories, where monopole operator R-charge can be computed semiclassically. It is conceivable that the $\grU(1)_R$ charge does not change under the RG flow, which would then determine it in the superconformal theory. \\

As we have noted, in some quiver theories the induced R-charge of the monopole in the UV is simply proportional to the sum over the R-charges of all the fermions. If a ``parent'' quiver Yang-Mills gauge theory can be written down in 4-d with the same superpotential, then all the fermion R-charges in it are the same as in 3-d. In this ``parent theory'' the sum over all fermion R-charges determines the $\grU(1)_R$ anomaly. In particular, if this quantity vanishes, then the 4-d gauge theory is superconformal. Thus, it is tempting to conjecture a relation between $\grU(1)_R$ anomaly in a parent 4-d gauge theory and the induced monopole operator R-charge in a descendant 3-d gauge theory. For example, in the class of $\grU(N)\times \grU(N)$ theories we have considered in this paper, the induced monopole R-charge is $\sim (1- N_f/2)$. In its parent 4-d gauge theory, the $\grU(1)_R$ anomaly coefficient is $(1- N_f/2)N$, with the first term due to an adjoint gluino of R-charge 1, and the second due to $N_f$ bifundamentals of R-charge $-1/2$. The anomaly cancellation for $N_f=2$ singles out the 4-d superconformal gauge theory describing D3-branes on the conifold \cite{Klebanov:1998hh}. \\

This discussion suggests that, if a quiver gauge theory is superconformal in 4-d, then at least some monopole operators in its 3-d descendant (presumably the ones that correspond to turning on monopoles in all gauge groups) have vanishing monopole operator dimensions. Clearly, the possibility of a connection between monopole R-charges in 3-d and anomaly coefficients in 4-d requires a more detailed study.

%%%%%%%%%%%%%%%%%%%%%%%%%%%%%%%%%%%%%%%%%%%%%%%%%%%%%%%%%%%%%%%%%%%%%%%%%%%
%%%%%%%%%%%%%%%%%%%%%%%%%%%%%%%%%%%%%%%%%%%%%%%%%%%%%%%%%%%%%%%%%%%%%%%%%%%
\section*{Acknowledgments}

We are very grateful to Chris Herzog, Daniel Jafferis, Arvind Murugan, Silviu Pufu, Mikael Smedb\" ack, Edward Witten, and especially Juan Maldacena for many interesting and helpful discussions. M.K.B.\ would like to express his gratitude for hospitality to LPTHE (Jussieu, Paris), and I.R.K.\ to the Galileo Galilei Institute (Florence), where parts of this work were carried out. This research is supported in part by the National Science Foundation Grant No.~PHY-0756966.

%%%%%%%%%%%%%%%%%%%%%%%%%%%%%%%%%%%%%%%%%%%%%%%%%%%%%%%%%%%%%%%%%%%%%%%%%%%
%%%%%%%%%%%%%%%%%%%%%%%%%%%%%%%%%%%%%%%%%%%%%%%%%%%%%%%%%%%%%%%%%%%%%%%%%%%
\appendix

%%%%%%%%%%%%%%%%%%%%%%%%%%%%%%%%%%%%%%%%%%%%%%%%%%%%%%%%%%%%%%%%%%%%%%%%%%%
%%%%%%%%%%%%%%%%%%%%%%%%%%%%%%%%%%%%%%%%%%%%%%%%%%%%%%%%%%%%%%%%%%%%%%%%%%%
\section{Notation and conventions}
\label{sec:notation}

The main part of our computations are performed on $\Reals\times\Sphere^2$ with metric $ds^2 = g_{mn}dx^m dx^n = d\tau^2 + d\theta^2 + \sin^2\theta\,d\varphi^2$. As Dirac matrices in the tangent frame we use $(\gamma^a)_\alpha{}^\beta = (-\sigma^2,\sigma^1,\sigma^3)$, which satisfy $\gamma^a \gamma^b = \delta^{ab} + i\levi^{abc} \gamma^c$. Spinor indices are raised and lowered from the left, $\psi^\alpha = \levi^{\alpha\beta} \psi_\beta$ and $\psi_\alpha = \levi_{\alpha\beta} \psi^\beta$, with $\levi^{12} = -\levi_{12} = 1$. Note that $(\gamma^a)_{\alpha\beta} = (-i\unit,-\sigma^3,\sigma^1)$ are symmetric and we also have $(\gamma^a)_\alpha{}^\beta = (\gamma^a)^\beta{}_\alpha$. For contracting spinor indices we use the NW-SE convention, e.g $\psi \gamma^m \gamma^n \chi \equiv \psi^\alpha (\gamma^m)_\alpha{}^\beta (\gamma^n)_\beta{}^\gamma \chi_\gamma$ etc. The spin connection is
\be
  \nabla_m \psi = (\partial_m + \omega_m) \psi
  \comma
  \omega_m = \quarter \omega_{m a b} \gamma^{ab}
  \comma
  \gamma^{ab} = \half \comm{ \gamma^a }{ \gamma^b }
\ee
with the only non-zero component being $\omega_{\varphi 21} = -\omega_{\varphi 12} = \cos\theta$. For raising and lowering $\grSU(2)_R$ indices (also called $a,b,\ldots$) we use the same conventions as for spinor indices. The standard index position for Pauli matrices is $(\sigma_i)_a{}^b$.

The component expansion of the $\superN=2$ superfields are given as follows. The vector superfields containing the $\grU(N)\times\hat{\grU}(N)$ gauge fields are
\be
  \VV \eq 2i \, \theta \bar{\theta} \, \sigma(x)
    - 2 \, \theta\gamma^m\bar{\theta} \, A_m(x)
    + \sqrt{2} i \, \theta^2 \, \bar{\theta} \chi_\sigma^\dagger(x)
    - \sqrt{2} i \, \bar{\theta}^2 \, \theta \chi_\sigma(x)
    + \theta^2 \, \bar{\theta}^2 \, \auxD(x)
  \; , \\
  \hat{\VV} \eq
     2i \, \theta \bar{\theta} \, \hat{\sigma}(x)
    - 2 \, \theta\gamma^m\bar{\theta} \, \hat{A}_m(x)
    + \sqrt{2} i \, \theta^2 \, \bar{\theta} \hat{\chi}_\sigma^\dagger(x)
    - \sqrt{2} i \, \bar{\theta}^2 \, \theta \hat{\chi}_\sigma(x)
    + \theta^2 \, \bar{\theta}^2 \, \hat{\auxD}(x)
    \; ,
\ee
the chiral superfields in adjoints of the two gauge group factors are
\begin{align}
  \Phi             & = \phi(x_L)               + \sqrt{2} \, \theta       \chiphi(x_L)                 + \theta^2       \, F_\phi(x_L) \; , &
  \bar{\Phi}       & = \phi^\dagger(x_R)       - \sqrt{2} \, \bar{\theta} \chi^\dagger_\phi(x_R)       - \bar{\theta}^2 \, F^\dagger_\phi(x_R) \; , \\
  \hat{\Phi}       & = \hat{\phi}(x_L        ) + \sqrt{2} \, \theta       \hchiphi(x_L)                + \theta^2       \, \hat{F}_\phi(x_L) \; , &
  \hat{\bar{\Phi}} & = \hat{\phi}^\dagger(x_R) - \sqrt{2} \, \bar{\theta} \hat{\chi}^\dagger_\phi(x_R) - \bar{\theta}^2 \, \hat{F}^\dagger_\phi(x_R) \; ,
\end{align}
and the bifundamentals matter fields are
\begin{align}
  \ZZ       & = Z(x_L)         + \sqrt{2} \, \theta\zeta(x_L)                + \theta^2       \, F(x_L) \; , &
  \bar{\ZZ} & = Z^\dagger(x_R) - \sqrt{2} \, \bar{\theta}\zeta^\dagger(x_R)  - \bar{\theta}^2 \, F^\dagger(x_R) \; , \\
  \WW       & = W(x_L)         + \sqrt{2} \, \theta\omega(x_L)               + \theta^2       \, G(x_L) \; , &
  \bar{\WW} & = W^\dagger(x_R) - \sqrt{2} \, \bar{\theta}\omega^\dagger(x_R) - \bar{\theta}^2 \, G^\dagger(x_R) \; ,
\end{align}
where $x_L^m = x^m - i \theta \gamma^m \bar{\theta}$ and $x_R^m = x^m + i \theta \gamma^m \bar{\theta}$.

%%%%%%%%%%%%%%%%%%%%%%%%%%%%%%%%%%%%%%%%%%%%%%%%%%%%%%%%%%%%%%%%%%%%%%%%%%%
%%%%%%%%%%%%%%%%%%%%%%%%%%%%%%%%%%%%%%%%%%%%%%%%%%%%%%%%%%%%%%%%%%%%%%%%%%%
\section{$\superN=3$ Chern-Simons Yang-Mills on $\Reals^{1,2}$}
\label{sec:N3-YM-CS-R12}

We present the action of $\superN=3$ Chern-Simons Yang-Mills theory on $\Reals^{1,2}$ with signature $(-,+,+)$. For further explanations see \secref{sec:CSYM}. The kinetic terms are given by
\be
  \Action_{\mathrm{kin}} \eq \int\!d^3x\: \tr \Bigsbrk{
   - \tfrac{1}{2g^2} F^{\mu\nu} F_{\mu\nu}
   + \kappa \, \levi^{\mu\nu\lambda} \bigbrk{
            A_\mu \partial_\nu A_\lambda
          + \tfrac{2i}{3} A_\mu A_\nu A_\lambda }
   \nl\hspace{17mm}
   - \tfrac{1}{2g^2} \hat{F}^{\mu\nu} \hat{F}_{\mu\nu}
   - \kappa \, \levi^{\mu\nu\lambda} \bigbrk{
              \hat{A}_\mu \partial_\nu \hat{A}_\lambda
            + \tfrac{2i}{3} \hat{A}_\mu \hat{A}_\nu \hat{A}_\lambda }
   \nl[1mm]\hspace{17mm}
   - \deriD_\mu X^\dagger \deriD^\mu X
   + i \xi^\dagger \slashed{\deriD} \xi
   \nl[1mm]\hspace{17mm}
   - \tfrac{1}{2g^2} \deriD_\mu \phi^a_b \deriD^\mu \phi^b_a
   - \half \kappa^2 g^2 \, \phi^a_b \phi^b_a
   - \tfrac{1}{2g^2} \deriD_\mu \hat{\phi}^a_b \deriD^\mu \hat{\phi}^b_a
   - \half \kappa^2 g^2 \, \hat{\phi}^a_b \hat{\phi}^b_a
   \nl[1mm]\hspace{17mm}
   - \tfrac{i}{2g^2} \lambda^{ab} \slashed{\deriD} \lambda_{ab}
   - \tfrac{\kappa}{2} \, i \lambda^{ab} \lambda_{ba}
   - \tfrac{i}{2g^2} \hat{\lambda}^{ab} \slashed{\deriD} \hat{\lambda}_{ab}
   + \tfrac{\kappa}{2} \, i \hat{\lambda}^{ab} \hat{\lambda}_{ba}
  }
\ee
and the interactions by
\be
  \Action_{\mathrm{int}} \eq \int\!d^3x\: \tr \Bigsbrk{
   - \kappa g^2 \, X^\dagger_a \phi^a_b X^b
   + \kappa g^2 \, X^a \hat{\phi}^b_a X^\dagger_b
   - i \xi^\dagger_a \phi^a_b \xi^b
   - i \xi^a \hat{\phi}^b_a \xi^\dagger_b
   \nl[-2mm]\hspace{17mm}
   + \levi_{ac} \lambda^{cb} X^a \xi^\dagger_b
   - \levi^{ac} \lambda_{cb} \xi^b X^\dagger_a
   - \levi_{ac} \hat{\lambda}^{cb} \xi^\dagger_b X^a
   + \levi^{ac} \hat{\lambda}_{cb} X^\dagger_a \xi^b
   \nl[1mm]\hspace{17mm}
   + \tfrac{\kappa}{6} \phi^a_b \comm{\phi^b_c}{\phi^c_a}
   + \tfrac{\kappa}{6} \hat{\phi}^a_b \comm{\hat{\phi}^b_c}{\hat{\phi}^c_a}
   - \tfrac{1}{2g^2} i \lambda_{ab} \comm{\phi^b_c}{\lambda^{ac}}
   + \tfrac{1}{2g^2} i \hat{\lambda}_{ab} \comm{\hat{\phi}^b_c}{\hat{\lambda}^{ac}}
   \nl[1mm]\hspace{17mm}
   - \tfrac{g^2}{4} (X \sigma_i X^\dagger) (X \sigma_i X^\dagger)
   - \tfrac{g^2}{4} (X^\dagger \sigma_i X) (X^\dagger \sigma_i X)
   \nl[1mm]\hspace{17mm}
   - \tfrac{1}{2} (X X^\dagger) \phi^a_b \phi^b_a
   - \tfrac{1}{2} (X^\dagger X) \hat{\phi}^a_b \hat{\phi}^b_a
   - X^\dagger_{Aa} \phi^b_c X^{Aa} \hat{\phi}^c_b
   \nl\hspace{17mm}
   + \tfrac{1}{8g^2} \comm{\phi^a_b}{\phi^c_d} \comm{\phi^b_a}{\phi^d_c}
   + \tfrac{1}{8g^2} \comm{\hat{\phi}^a_b}{\hat{\phi}^c_d} \comm{\hat{\phi}^b_a}{\hat{\phi}^d_c}
  } \; .
\ee
The supersymmetry variations with parameter $\eps_{ab} = \eps_i (\sigma_i)_{ab}$ in the $\rep{3}$ of $\grSU(2)_R$ read
\be
  \delta A_\mu            \eq -\ihalf \eps_{ab} \gamma_\mu \lambda^{ab} \; , \\
  \delta \lambda^{ab} \eq \half \levi^{\mu\nu\lambda} F_{\mu\nu} \gamma_\lambda \eps^{ab}
                              - i \slashed{\deriD} \phi^b_c \eps^{ac}
                              + \tfrac{i}{2} \comm{\phi^b_c}{\phi^c_d} \eps^{ad}
                              + \kappa g^2 \, i \phi^b_c \eps^{ac}
                              + g^2 \, i X^a X^\dagger_c \eps^{cb}
                              - \tfrac{ig^2}{2} (X X^\dagger) \eps^{ab} \; , \nn \\
  \delta \phi^a_b     \eq - \eps_{cb} \lambda^{ca}
                              + \half \delta^a_b \eps_{cd} \lambda^{cd} \; , \nn
\ee
\be
  \delta \hat{A}_\mu        \eq -\ihalf \eps_{ab} \gamma_\mu \hat{\lambda}^{ab} \; , \\
  \delta \hat{\lambda}^{ab} \eq \half \levi^{\mu\nu\lambda} \hat{F}_{\mu\nu} \gamma_\lambda \eps^{ab}
                                + i \slashed{\deriD} \hat{\phi}^b_c \eps^{ac}
                                + \tfrac{i}{2} \comm{\hat{\phi}^b_c}{\hat{\phi}^c_d} \eps^{ad}
                                + \kappa g^2 \, i \hat{\phi}^b_c \eps^{ac}
                                - g^2 \, i \eps^{bc} X^\dagger_c X^a
                                + \tfrac{ig^2}{2} (X^\dagger X) \eps^{ab} \; , \nn \\
  \delta \hat{\phi}^a_b     \eq - \eps_{cb} \hat{\lambda}^{ca}
                                + \half \delta^a_b \eps_{cd} \hat{\lambda}^{cd} \; , \nn
\ee
\begin{align}
  \delta X^{Aa}           & = - i \eps^a_b \xi^{Ab} \; , &
  \delta \xi^{Aa}         & = \slashed{\deriD} X^{Ab} \eps^a_b
                              + \phi^a_b \eps^b_c X^{Ac}
                              + X^{Ac} \eps^b_c \hat{\phi}^a_b \; , \\
  \delta X^\dagger_{Aa}   & = - i \xi^\dagger_{Ab} \eps^b_a \; , &
  \delta \xi^\dagger_{Aa} & = \slashed{\deriD} X^\dagger_{Ab} \eps^b_a
                              + \hat{\phi}^b_a \eps^c_b X^\dagger_{Ac}
                              + X^\dagger_{Ac} \eps^c_b \phi^b_a \; . \nn
\end{align}

%%%%%%%%%%%%%%%%%%%%%%%%%%%%%%%%%%%%%%%%%%%%%%%%%%%%%%%%%%%%%%%%%%%%%%%%%%%
%%%%%%%%%%%%%%%%%%%%%%%%%%%%%%%%%%%%%%%%%%%%%%%%%%%%%%%%%%%%%%%%%%%%%%%%%%%
\section{Monopole spinor harmonics}
\label{sec:monopole-spinor-harmonics}

We define monopole spinor harmonics as eigenspinors of the Dirac operator on the sphere in a monopole background with magnetic charge $q$:
\be
  -i \slashed{\deriD}_\Sphere \Bsi^\pm_{qjm} \eq \Delta^\pm_{jq} \Bsi^\pm_{qjm}
\ee
with eigenvalues
\be
  \Delta^\pm_{jq} = \pm \half \sqrt{(2j+1)^2 - q^2}
\ee
for $j=\frac{\abs{q}-1}{2},\frac{\abs{q}+1}{2},\ldots$ and $m = -j,-j+1,\ldots,j$. The spectrum is drawn in \figref{fig:ev-Dirac} on page \pageref{fig:ev-Dirac}. These spinors also satisfy
\be \label{eqn:msh-gammatau}
  \gamma^\tau \, \Bsi^{\pm}_{qjm} = \Bsi^{\mp}_{qjm} \; ,
\ee
which couples modes with positive and negative eigenvalue. The lowest modes, $j=\frac{\abs{q}-1}{2}$, which only exists for $q\not=0$ are zero-modes and the corresponding $\Bsi^\pm$-spinors are not independent. We introduce a special notation for them
\be
  \Bsi^0_{qm} \equiv \frac{1}{\sqrt{2}} \Bigbrk{ \Bsi^+_{qjm} + \sign(q) \, \Bsi^-_{qjm} }_{j=\frac{\abs{q}-1}{2}} \; .
\ee
Then \eqref{eqn:msh-gammatau} implies
\be \label{eqn:msh-gammatau-zero}
  \gamma^\tau \Bsi^0_{qm} = \sign(q) \, \Bsi^0_{qm} \; .
\ee
Further properties are the orthogonality
\be \label{eqn:msh-orthogonal}
  \int \!d\Omega\: \Bsi^{0 \dagger}_{qm} \, \gamma^\tau \, \Bsi^{0}_{qm'} = i \delta_{mm'}
  \comma
  \int \!d\Omega\: \Bsi^{\eps \dagger}_{qjm} \, \gamma^\tau \, \Bsi^{\eps'}_{qj'm'} = i \delta^{\eps\eps'} \, \delta_{jj'} \, \delta_{mm'} \; ,
\ee
and completeness relations
\be
  \sum_m \Bsi^0_{qm}(\Omega) \Bsi^{0\dagger}_{qm}(\Omega') + \sum_{jm\eps} \Bsi^\eps_{qjm}(\Omega) \Bsi^{\eps\dagger}_{qjm}(\Omega') = i\gamma^\tau \delta^{2}(\Omega-\Omega') \; .
\ee

We also note the explicit expressions. The generalization of the spinor harmonics in \cite{Abrikosov:2002jr} to non-zero monopole background is
\be
  \tilde{\Bsi}^+_{qjm} \eq \sqrt{\tfrac{1+r_{qj}}{2}} \, \Omega^+_{qjm} + i\sign(q) \, \sqrt{\tfrac{1-r_{qj}}{2}} \, \Omega^-_{qjm} \; , \\
  \tilde{\Bsi}^-_{qjm} \eq \sign(q) \sqrt{\tfrac{1-r_{qj}}{2}} \, \Omega^+_{qjm} + i \sqrt{\tfrac{1+r_{qj}}{2}} \, \Omega^-_{qjm} \; ,
\ee
with $r_{qj} = \sqrt{1-\frac{q^2}{(2j+1)^2}}$ and
\be
  \Omega^\pm_{qjm} \eq \frac{(-)^{j-m}  \lrbrk{\frac{i}{2}}^{j+\half} \lrbrk{j+\half}}{\sqrt{\Gamma\lrbrk{j+\tfrac{3}{2}-\tfrac{q}{2}} \Gamma\lrbrk{j+\tfrac{3}{2}+\tfrac{q}{2}}}}
  \, \sqrt{\frac{(j-m)!}{(j+m)!}}
  \, \frac{e^{i\lrbrk{m+\tfrac{q}{2}}\varphi}}{\sqrt{2\pi}} \times \\ \nn && \times
  \matr{c}{
  \mp\sqrt{\mp i} \, (1-x)^{m_{-+}/2}
                  \, (1+x)^{m_{+-}/2}
                  \, {\displaystyle \frac{d^{j+m}}{dx^{j+m}}} \, \Bigbrk{ (1-x)^{j_{+-}} (1+x)^{j_{-+}} } \\[4mm]
  \pm\sqrt{\pm i} \, (1-x)^{m_{++}/2}
                  \, (1+x)^{m_{--}/2}
                  \, {\displaystyle \frac{d^{j+m}}{dx^{j+m}}} \, \Bigbrk{ (1-x)^{j_{--}} (1+x)^{j_{++}} }
  }
\ee
where
\be
  j_{\eps_1\eps_2} \equiv j + \eps_1 \tfrac{1}{2} + \eps_2 \tfrac{q}{2}
  \comma
  m_{\eps_1\eps_2} \equiv m + \eps_1 \tfrac{1}{2} + \eps_2 \tfrac{q}{2}
  \; .
\ee
We rotate to our basis of Dirac matrices by defining $\Bsi^\pm_{qjm} = \tfrac{1}{\sqrt{2}}(1-i\sigma^1) \tilde{\Bsi}^\pm_{qjm}$.

%%%%%%%%%%%%%%%%%%%%%%%%%%%%%%%%%%%%%%%%%%%%%%%%%%%%%%%%%%%%%%%%%%%%%%%%%%%
%%%%%%%%%%%%%%%%%%%%%%%%%%%%%%%%%%%%%%%%%%%%%%%%%%%%%%%%%%%%%%%%%%%%%%%%%%%
\bibliographystyle{nb}
\bibliography{monopoles}

\end{document}